\documentclass[prl,showpacs,twocolumn,
               amsmath,amssymb,superscriptaddress,
               floatfix,nofootinbib]{revtex4-1}

\usepackage[utf8]{inputenc}
\usepackage{amsmath,amssymb,amsbsy,amsfonts}
\usepackage{bm}
\usepackage{times}

\usepackage{graphicx}
\usepackage{float}
\usepackage{morefloats}
\usepackage{rotating}

\usepackage[caption=false]{subfig}

\usepackage{booktabs}
\usepackage{multirow}
\usepackage{tabularx}
\usepackage{adjustbox}
\usepackage{nicefrac}
\usepackage{slashed}
\usepackage{verbatim}
\usepackage[dvipsnames]{xcolor}
\usepackage{ragged2e}

\usepackage{hyperref}

\begin{document}

\title{Two-pion exchange contributions to the relativistic chiral nuclear force at N$^3$LO}

\author{Jun-Xu Lu}
\affiliation{School of Physics, Beihang University, Beijing 102206, China}

\author{Li-Sheng Geng}
\email[Corresponding author: ]{lisheng.geng@buaa.edu.cn}
\affiliation{Sino-French Carbon Neutrality Research Center, \'Ecole Centrale de P\'ekin/School of General Engineering, Beihang University, Beijing 100191, China}
\affiliation{School of Physics, Beihang University, Beijing 102206, China}
\affiliation{Peng Huanwu Collaborative Center for Research and Education, Beihang University, Beijing 100191, China}
\affiliation{Beijing Key Laboratory of Advanced Nuclear Materials and Physics, Beihang University, Beijing 102206, China }
\affiliation{Southern Center for Nuclear-Science Theory (SCNT), Institute of Modern Physics, Chinese Academy of Sciences, Huizhou 516000, China}

\begin{abstract}
We present the two-pion exchange contributions to the nucleon-nucleon interaction up to next-to-next-to-next-to leading order (N$^3$LO) in covariant baryon chiral perturbation theory. Both one-loop and two-loop diagrams are calculated with the spectral functional regularization. We show that the phase shifts for partial waves with total angular momentum $3\le J\le 5$ are in better agreement with the partial wave analysis from the Nijmegen or the SAID group than their N$^2$LO counterparts. In addition, the relativistic chiral force exhibits better convergence than its non-relativistic counterpart, suggesting the importance of relativistic corrections. 
\end{abstract}

\maketitle

\section{Introduction}

Chiral nuclear forces have attracted extensive attention since Weinberg initiated applying chiral effective field theory (ChEFT) to the construction of nucleon-nucleon (NN) interactions in the 1990s~\cite{Weinberg:1990rz,Weinberg:1991um,Weinberg:1992yk}. At present, the fifth order of chiral expansion~\cite{Epelbaum:2014sza,Reinert:2017usi,Entem:2017gor} and the dominant sixth order contribution~\cite{RodriguezEntem:2020jgp} have been accomplished in the non-relativistic (heavy baryon) scheme of ChEFT, the precision of which has reached the level of the most refined phenomenological forces, such as Argonne $\textrm{V}_{18}$~\cite{Wiringa:1994wb} and CD-Bonn~\cite{Machleidt:2000ge}. Weinberg’s chiral nuclear force has proven to be a great success and has become the standard microscopic input for ab initio nuclear calculations. For details and recent progress, we refer to Refs.~\cite{Epelbaum:2008ga,Machleidt:2011zz,Machleidt:2024bwl} and references therein.

However, the Weinberg chiral nuclear force is constructed within the framework of heavy-baryon chiral effective field theory (HBChEFT), wherein the non-relativistic expansion applied to baryon propagators and Dirac spinors results in a relatively slow convergence, as dictated by the power counting rules based on Weinberg’s naive dimensional analysis. It was argued that an accurate description of the neutron-deuteron scattering requires at least the next-to-next-to-next-to-next-to-leading order (N$^4$LO) chiral nuclear force~\cite{Epelbaum:2019zqc}.

Meanwhile, it has been noted that Lorentz covariance significantly improves convergence in chiral effective field theories. This conclusion is corroborated by numerous studies based on covariant ChEFT within the single-baryon system, including investigations into baryon masses ~\cite{Ren:2012aj,Ren:2013oaa,Ren:2013wxa,Ren:2013dzt,Ren:2016aeo,Chen:2024twu,Liang:2025cjd}, sigma terms~\cite{Ren:2014vea,Ren:2017fbv,Liang:2025adz}, magnetic moments ~\cite{Geng:2008mf,Liu:2018euh,Shi:2018rhk,Xiao:2018rvd,Shi:2021kmm}, $\pi N$ scattering~\cite{Chen:2012nx,Siemens:2016hdi,Siemens:2016jwj,Lu:2018zof}, Generalized polarizabilities of the nucleon~\cite{Hagelstein:2015egb,Lensky:2016nui}, octet-baryon axial vector charges~\cite{Ledwig:2014rfa}, weak pion production off the nucleon~\cite{Yao:2018pzc,Yao:2019avf}, and hyperon weak radiative decays~\cite{Shi:2022dhw}. For an early short review, see Ref.~\cite{Geng:2013xn}.

Most recently, we proposed a relativistic framework for constructing nuclear forces within covariant baryon chiral effective field theory~\cite{Ren:2016jna}. It retains the full Dirac spinors and Clifford algebra in the covariant chiral effective Lagrangian. It solves the relativistic scattering equation rather than the Lippmann-Schwinger equation to account for non-perturbative effects. Recently, such a relativistic chiral nuclear force was constructed up to next-to-next-to-leading order (N$^2$LO)~\cite{Xiao:2018jot,Xiao:2020ozd,Wang:2021kos,Lu:2021gsb}. The descriptions of the phase shifts up to $T_{\text{lab}}=200$ MeV are in good agreement with the Nijmegen partial-wave analysis results~\cite{Stoks:1993tb} and are also compatible with the results of the Weinberg N$^3$LO chiral forces~\cite{Entem:2003ft,Machleidt:2011zz,Epelbaum:2014efa,Epelbaum:2014sza}, which again demonstrates a satisfactory convergence. For a short review on this topic, we refer to Ref.~\cite{Lu:2025syk}.

A critical question always associated with the application of ChEFT concerns the order to which the chiral expansion ought to be truncated. This question is closely related to the convergence of power-counting rules, the accuracy with which observables are described, and the estimation of theoretical uncertainties. On the one hand, it is necessary to verify that the contribution of the next higher order does not degrade the description of phase shifts provided by the N$^2$LO relativistic chiral nuclear force. On the other hand, the accuracy of the N$^2$LO relativistic chiral nuclear force is not yet fully comparable to that of the state-of-the-art N$^4$LO$+$ non-relativistic chiral nuclear force~\cite{Reinert:2017usi}.

Therefore, to confirm the convergence pattern of our relativistic chiral nuclear force and, what is more important, to further improve the accuracy, we extend the calculation one order higher to N$^3$LO within the covariant chiral effective field theory. In the present manuscript, we focus on the peripheral partial waves with total angular momentum $J \ge 3$. For these partial waves, except the $^3D_3$ partial waves, non-perturbative effects are less pronounced, and perturbative treatments are adequate.

The manuscript is organized as follows. In Sec.~\ref{sec1}, we briefly introduce the formalism of calculating two-loop diagrams. Specifically, the relevant covariant $\pi N$ Lagrangians are shown explicitly in Sec .~\ref {subsec:Lag}, followed by a detailed introduction on how we apply the spectral function method to calculate two-loop diagrams in Sec .~\ref {subsec:SFR}. Then, in Sec.~\ref{sec:results}, we present the phase shifts and mixing angles for these peripheral partial waves, and compare them with results obtained via the dimensional regularization method. Finally, the summary is given in Sec.~\ref{sec:Sum}.

\section{Formalism}
\label{sec1}

The Feynman diagrams to be calculated up to N$^3$LO can be specified by the following power-counting rules of ChEFT. The chiral order $\nu$ for a certain Feynman diagram with $L$ loops is counted as
\begin{equation}\label{eq:pcr1}  
   \nu = 4 L-2N_\pi-N_N+\sum_k kV_k 
\end{equation}
where $N_{\pi,N}$ are the numbers of $\pi$ and nucleon propagators, and $V_k$ is the number of $k$-th order vertices. Thus, up to N$^3$LO, i.e., $\nu=4$, the potential $V$ contains the following parts involving pions,
\begin{align}\label{eq:pcr2}  
   V^{\text{total}} & = V^{(0)} + V^{(2)} + V^{(3)} + V^{(4)},
\end{align}
where the superscripts denote the chiral orders of each piece. Each of these components can be further subdivided according to the number of exchanged pions,
\begin{align}\label{eq:pcr3}  
   V^{(0)} &= V_{1\pi}^{(0)}, \nonumber\\
   V^{(2)} &= V_{1\pi}^{(2)} + V_{2\pi}^{(2)}, \nonumber\\
   V^{(3)} &= V_{1\pi}^{(3)} + V_{2\pi}^{(3)},\nonumber\\
   V^{(4)} &= V_{1\pi}^{(4)} + V_{2\pi}^{(4)}  + V_{3\pi}^{(4)}.
\end{align}
At leading order (LO), only one-pion exchange (OPE) contributes, characterizing the long-range part, while at next-to-leading order (NLO) and N$^2$LO, leading and sub-leading two-pion exchanges (TPE), which represent the intermediate-range component, arise. At N$^3$LO, novel three-pion exchange diagrams appear. Besides, the TPE contributions include novel two-loop diagrams.

The explicit expression for the LO OPE contribution in the relativistic framework is given in Ref.~\cite{Ren:2016jna}. The higher order corrections to the OPE involving loops and counterterms are taken into account by taking $g_A=1.290$ instead of $g_A=1.276$~\cite{UCNA:2010les}, as was done in the non-relativistic framework~\cite{Machleidt:2011zz}. 

On the other hand, the three-pion exchange contributions at N$^3$LO have already been calculated by the Munich group~\cite{Kaiser:1999ff,Kaiser:1999jg} in the non-relativistic frame and were found to be negligible. It was argued that the three-pion exchanges only involve leading-order $\pi N$ vertices, which are known to be weaker than the diagrams containing NLO $\pi N$ vertices. The fact that the subleading TPE contributions incorporating NLO $\pi N$ vertices are much larger than the leading TPE part containing only LO $\pi N$ vertices supports such a conclusion. In the relativistic TPE contributions, we found results similar to those depicted in Ref.~\cite{Xiao:2020ozd}. Therefore, following the non-relativistic chiral nuclear force studies, we tentatively neglect the three-pion exchanges for the time being and concentrate on the dominant TPE contributions up to N$^3$LO. 


\begin{figure*}[htpb]
    \centering
    \includegraphics[width=0.7\linewidth]{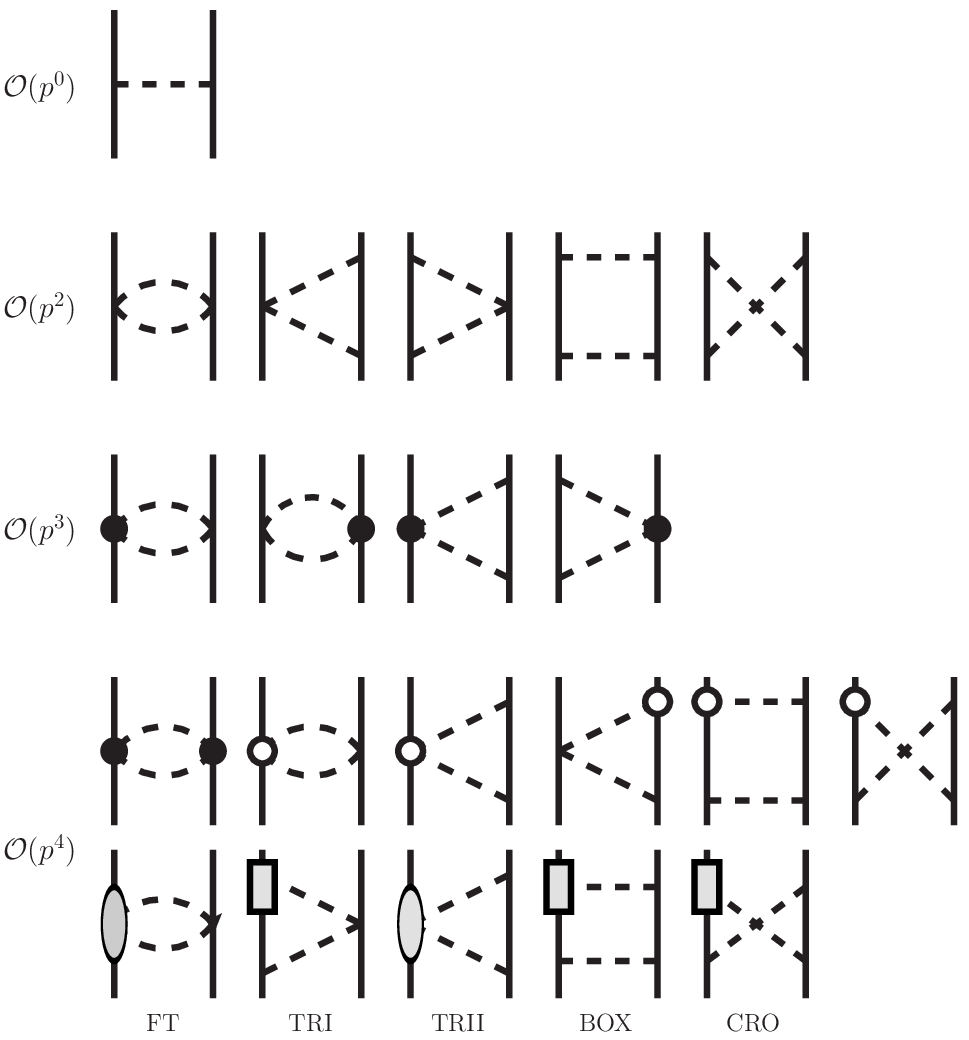}
    \caption{Pion exchange contributions at various orders are shown. The solid and open dots denote the vertices of $\mathcal{L}_{\pi N}^{(2)}$ in Eq.~(\ref{eq:lagpiN2}) and $\mathcal{L}_{\pi N}^{(3)}$ in Eq.~(\ref{eq:lagpiN3}), respectively. The gray circles and boxes denote the corresponding one-loop level $\pi\pi$NN and $\pi NN$ interactions depicted in Fig.~\ref{fig:N3LO-part1} and Fig.~\ref{fig:N3LO-part2}. All diagrams that occur at each channel are shown, except for the N$^3$LO($\mathcal{O}(p^4)$) part, where the left-right exchanged diagrams and time-reversal symmetric diagrams are not explicitly displayed but included in the calculation.  }
    \label{fig:Feynman-All}
\end{figure*}

\subsection{Lagrangians}\label{subsec:Lag}

To calculate the pion exchange contributions, one needs the following effective Lagrangians consisting of $\pi N$ and purely mesonic Lagrangians up to N$^2$LO ($\mathcal{O}(p^3)$), which read~\cite{Fettes:2000gb,Chen:2012nx}
\begin{equation}\label{eq:lag1}  
   \mathcal{L}_{\text{eff}}=\mathcal{L}_{\pi N}^{(1)}+\mathcal{L}_{\pi N}^{(2)}+\mathcal{L}_{\pi N}^{(3)}+\mathcal{L}_{\pi \pi}^{(2)}+\mathcal{L}_{\pi \pi}^{(4)}
\end{equation}
where the superscripts again denote the chiral orders, respectively. The $\pi N$ Lagrangians at each order take the following form~\cite{Chen:2012nx}
\begin{equation}\label{eq:lagpiN1}  
   \mathcal{L}_{\pi N}^{(1)}=\bar{N}\left( i\slashed{D} -m +\frac{1}{2}g_A\slashed{u}\gamma^5\right)N,
\end{equation}
\begin{equation}\label{eq:lagpiN2} 
\begin{split}
   \mathcal{L}_{\pi N}^{(2)} &= c_{1}\langle \chi _{+}\rangle\bar{N}N - \frac{c_{2}}{4m^{2}}\langle u^{\mu }u^{\nu }\rangle \left( \bar{N}D_{\mu }D_{\nu }N + h.c.\right) \\
   &+ \frac{c_{3}}{2}\langle u^{\mu }u_{\mu }\rangle \bar{N}N - \frac{c_{4}}{4}\bar{N}\gamma ^{\mu }\gamma ^{\nu }\left[ u_{\mu },u_{\nu }\right] N,
\end{split}
\end{equation}
\begin{equation}\label{eq:lagpiN3} 
\begin{split}
   \mathcal{L}_{\pi N}^{(3)} &= \bar{N}\left\{-\frac{d_{1}+d_{2}}{4m}\left( \left[ u_{\mu },\left[ D_{\nu },u^{\mu }\right] + \left[ D^{\mu },u_{\nu }\right] \right] D^{\nu } + h.c.\right) \right.\\ 
&\left.+ \frac{d_{3}}{12m^{3}}\left( \left[ u_{\mu },\left[ D_{\nu },u_{\lambda }\right] \right] \right) \left( D^{\mu }D^{\nu }D^{\lambda } + sym.\right) + h.c. \right.\\
&\left.+ i\frac{d_{5}}{2m}\left( \left[ \chi_{-},u_{\mu }\right] D^{\mu } + h.c.\right) \right.\\
&\left.+ i\frac{d_{14}-d_{15}}{8m}\left( \sigma ^{\mu \nu }\left< \left[ D_{\lambda },u_{\mu }\right] u_{\nu } - u_{\mu }\left[ D_{\nu },u_{\lambda }\right] \right> D^{\lambda } + h.c.\right) \right.\\
&\left.+ \frac{d_{16}}{2}\gamma ^{\mu }\gamma ^{5}\left< \chi_{+}\right> u_{\mu } 
+ i\frac{d_{18}}{2}\gamma ^{\mu }\gamma ^{5}\left[ D_{\mu },\chi_{-}\right]
\right\} N.
\end{split}
\end{equation}
The covariant derivative $D_\mu$ is defined as $D_\mu=\partial_\mu+\Gamma_\mu$ with 
\begin{equation}\label{eq:lagpiN4}  
   \Gamma_\mu = \frac{1}{2}\left[ u^\dagger\partial_\mu u+u\partial_\mu u^\dagger    \right],
\end{equation}
and the axial current type quantity $u_\mu$ is defined as
\begin{align}\label{eq:lagpiN5}  
   u_\mu &= i\left[ u^\dagger\partial_\mu u-u\partial_\mu u^\dagger    \right], ~~~ u =\exp\left( \frac{i \Phi}{2f} \right),
\end{align}
where the pion field $\Phi$ takes the standard form
\begin{equation}\label{eq:lagpiN6}  
   \Phi = \begin{pmatrix} \pi^0 & \sqrt{2}\pi^+ \\ \sqrt{2}\pi^- & \pi^0 \end{pmatrix}.
\end{equation}
The remaining building blocks for the chiral symmetry-breaking part are 
\begin{equation}\label{eq:lagpiN7}  
   \chi_\pm=u^\dagger \chi u^\dagger\pm u\chi u,~~~~~\chi=\text{diag}\left( M^2,M^2 \right).
\end{equation}
Throughout the manuscript, we set $m_\pi=138$ MeV and $m=939$ MeV to denote the physical pion and nucleon masses, $f=92.4$ MeV to denote the pion decay constant, and $g_A=1.276$ to denote the physical value of the axial coupling constant~\footnote{As is explained below, we take the standard value $g_A=1.267$ for the renormalized $\pi N$ amplitude and $g_A=1.290$ for the rest parts. Actually, $g_A=1.267$ or $g_A=1.290$ yields almost the same results.}.

The $c_{1,2,3,4}$ and $d_{1,2,3,5,14,15,16,18}$ are the relevant low-energy constants (LECs), and we adopt the Fit I-$\mathcal{O}(p^3)$ in Ref.~\cite{Chen:2012nx} for their values, which are collected in Table~\ref {tab:LECpiN}.

\begin{table}[h]
\caption{LECs for the $\pi N$ Lagrangians up to $\mathcal{O}(p^3)$~\cite{Chen:2012nx}. The $c_i$ and $d_j$ have units of GeV$^{-1}$ and GeV$^{-2}$, respectively.}
\label{tab:LECpiN}
\centering
\setlength{\tabcolsep}{15pt}
\begin{tabular}{cccc}
 \hline\hline 
 \multicolumn{2}{c}{LEC($\mathcal{O}(p^2)$)} & \multicolumn{2}{c}{LEC($\mathcal{O}(p^3)$)}    \\
 \hline 
  $c_1$   & $-1.39$ & $d_1+d_2$ & $4.40$\\
  $c_2$   & $4.01$  & $d_3$  & $-3.02$\\
  $c_3$   & $-6.61$ & $d_5$ & $-0.62$\\
  $c_4$   & $3.92$  & $d_{14}-d_{15}$ & $-7.15$\\
          &         & $d_{18}$     &  $-0.56$ \\
 \hline 
\end{tabular}
\end{table}

The pure mesonic Lagrangians are also necessary for renormalization, and the relevant terms are given by
\begin{equation}
\begin{split}
   \mathcal{L}_{\pi\pi}^{(2)}=&\frac{f^2}{4}\langle u^\mu u_\mu+\chi_+ \rangle , \\
   \mathcal{L}_{\pi\pi}^{(4)}=&\frac{l_4}{8}\langle u^\mu u_\mu \rangle\langle \chi_+ \rangle + \frac{l_3+l_4}{16}\langle \chi_+ \rangle^2,
\end{split}
\end{equation}
where $l_3,l_4$ are the LECs for the $\pi\pi$ interactions, which, up to the chiral order of our interest, do not appear in the final amplitudes due to the renormalization. More details of the $\pi N$ interactions and the explicit corresponding expressions on the mass shell can be found in Ref.~\cite{Chen:2012nx}. 

\begin{figure*}[htpb]
    \centering
    \includegraphics[width=0.6\linewidth]{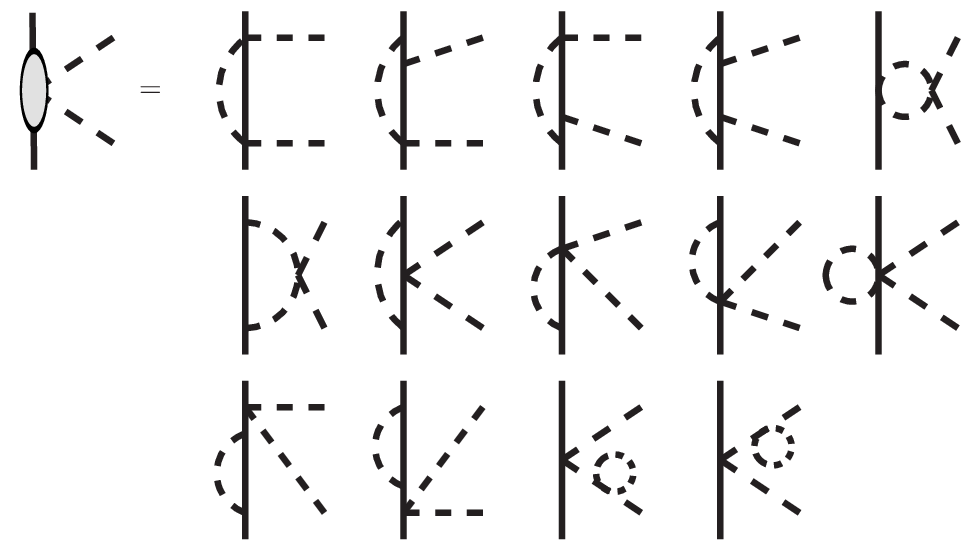}
    \caption{One-loop level $\pi N$ interactions of the Weinberg-Tomozawa-like terms. }
    \label{fig:N3LO-part1}
\end{figure*}

\begin{figure*}[htpb]
    \centering
    \includegraphics[width=0.6\linewidth]{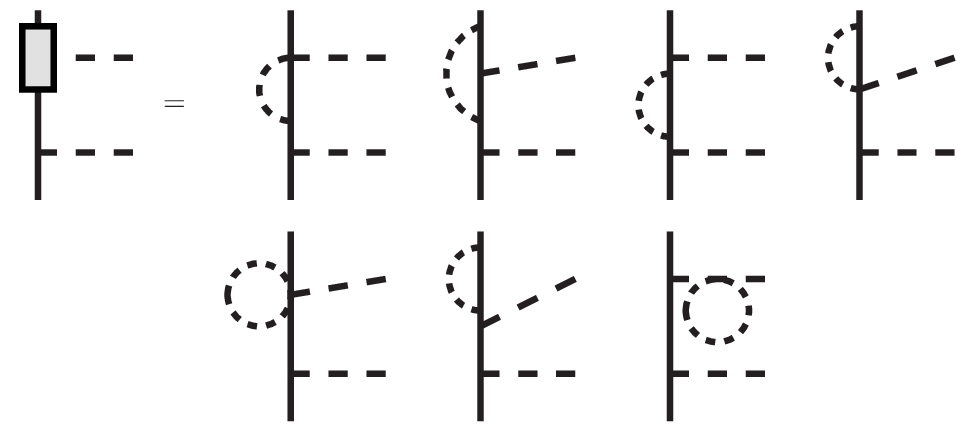}
    \caption{One-loop level $\pi N$  interactions of the Born-like terms. }
    \label{fig:N3LO-part2}
\end{figure*}

\subsection{spectral functional representation}\label{subsec:SFR}

Based on the power counting rules and the Lagrangians for $\pi N$ vertices, the Feynman diagrams for two-pion exchange contributions from NLO to N$^3$LO are depicted in Fig.~\ref{fig:Feynman-All}. At NLO, the TPEs are composed of football, triangle, planar box, and crossed box diagrams, each of which consists solely of the LO $\pi N$ vertices. At N$^2$LO, one of these LO $\pi N$ vertices is replaced by an NLO $\pi N$ vertex. Given the absence of an NLO $\pi NN$ vertex, only the football and triangle diagrams survive. Similarly, the N$^3$LO diagrams can be obtained via replacing the LO vertices with the N$^2$LO $\pi N$ ones. Since the N$^2$LO $\pi N$ amplitudes contain loops, the N$^3$LO TPE diagrams can be roughly categorized into several groups. The diagrams in the first row for N$^3$LO are obtained by replacing one corresponding LO $\pi N$ vertex with an N$^2$LO vertex or two LO $\pi N$ vertices with two corresponding NLO vertices, leading to the one-loop part. In contrast, the second row denotes those replacing the LO $\pi N$ vertex with one-loop corrections depicted as the shadowed circles or blocks in Fig.~\ref{fig:N3LO-part1} and Fig.~\ref{fig:N3LO-part2}, corresponding to the two-loop part. 

The calculation of multi-loop diagrams has always been a challenging task, especially when massive fermions are involved, since they generate extremely complex Lorentz structures due to Dirac algebra and axial couplings to Goldstone bosons. Currently, one strategy for handling dimensionally regularized Feynman integrals is as follows. One decomposes all Feynman integrals in a problem into a small set of basis functions, called master integrals (MIs), via integration-by-parts (IBP) reduction, and then evaluates these MIs. Most recently, an auxiliary mass flow method~\cite{Liu:2017jxz,Liu:2021wks}, which is also a differential equations method, was developed and implemented as the Mathematica package AMFlow~\cite{Liu:2022mfb,Liu:2022chg,Liu:2020kpc,Liu:2022tji}. This package provides an automatically high-precision numerical solution for dimensionally regularized Feynman integrals at arbitrary phase-space points, which has recently been applied to the calculation of the nucleon mass up to two-loop level($\mathcal{O}(p^5)$)~\cite{Liang:2025cjd} and to the extraction of the corresponding nucleon sigma term~\cite{Liang:2025adz}. 


However, in the present situation where nucleons are involved, the numerical solutions to dimensionally regularized Feynman integrals entail substantial computational time. This is especially true considering that the corresponding potential will also be used in non-perturbative calculations. At present, there is no precedent for calculating two-loop diagrams using the dimensional regularization method (DR) in the construction of nuclear forces or baryon-baryon interactions. Thus, we turn to another classic method, i.e., the so-called spectral function regularization method (SFR), to handle the extremely complicated Feynman integrals arising in the TPE contributions. It is equivalent to a cut-off regularization, and when the momentum cutoff $\Lambda$ is taken to infinity, it is identical to  dimensional regularization. Actually, such a regularization method has been widely applied in the construction of non-relativistic chiral nuclear forces, e.g., the most recently developed semi-local N$^4$LO$^+$ chiral force in momentum space~\cite{Reinert:2017usi} or the local N$^3$LO chiral force in coordinate space~\cite{Saha:2022oep}.

The basic idea of the spectral function regularization is that one first calculates the imaginary parts of the $NN$ amplitudes which result from the analytical continuation to time-like momentum transfer $q=i\mu-0^+$ with $\mu\ge 2M_\pi$. These imaginary parts are the mass-spectra (spectral functions) to be used in a continuous superposition of Yukawa functions which represent the $NN$ amplitudes themselves~\cite{Kaiser:1997mw,Chemtob:1972xt}, which reads
\begin{equation}\label{eq:SFR}  
   \mathcal{A}_{\text{NN}}=\frac{1}{\pi}\int_{4m_\pi^2}^\infty \text{d}\mu^2  \frac{\rho(\mu)}{\mu^2+q^2},
\end{equation}
where $\rho(\mu)$ is the corresponding mass spectra containing the whole dynamics corresponding to the exchanged $\pi\pi$ system. Note that since the spectral functions usually do not decrease for large $\mu$, which corresponds to ultra-divergences in dimensional regularization, a subtracted dispersion integral should be used. The LECs can absorb the related subtraction constants, so they do not introduce any additional ambiguity. 

Considering that EFTs only work within the low-$\mu$ expansion, we introduce an additional form factor $F(\mu)$ for the dispersion integrals as
\begin{equation}\label{eq:FormFactor}  
   F(\mu)=\exp(-\frac{\mu^4}{\Lambda^4}).
\end{equation}
Such a cutoff over the dispersion integral is related to another critical issue. The subleading TPE components governed by the NLO $\pi N$ LECs are found to be too large, leading to strong deviations from data for the higher energy region~\cite{Epelbaum:2003gr}. This deviation is attributed to the dimensional regularization used to calculate the subleading TPEs, which include the high-momentum components of exchanged pions that cannot be properly treated in an EFT. The introduction of the cutoff allows for the removal of these spurious short-distance physics associated with high-momentum intermediate states. It is demonstrated that this method can significantly accelerate the convergence of chiral expansions~\cite{Epelbaum:2003gr}.

We note that there are other choices for the form factor, such as the sharp cutoff $F(\mu)=\theta(\Lambda-\mu)$~\cite{Epelbaum:2003gr} or the local form $F(\mu)=e^{-\frac{\mu^2+q^2}{2\Lambda^2}}$~\cite{Reinert:2017usi}, which are specially designed not to induce any long-range finite cutoff artifacts. However, in the present work, we do not delve into the specific details of form factors; instead, we adopt a Gaussian form factor to ensure the spectral functions remain continuous as $\mu$ approaches infinity.

\begin{figure}[htpb]
    \centering
    \includegraphics[width=0.8\linewidth]{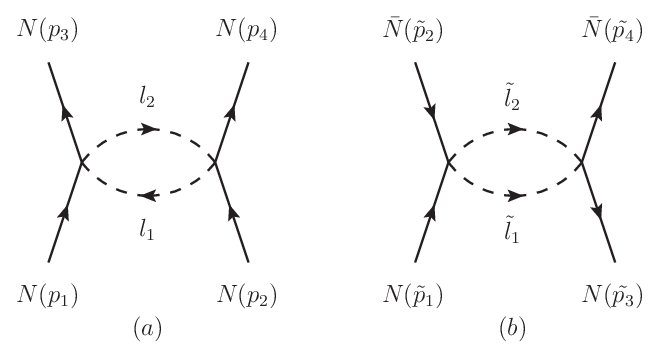}
    \caption{Football diagram of the LO TPE contributions (left). The right pattern is the corresponding $\bar{N}N\rightarrow \pi\pi \rightarrow \bar{N}N$ annihilation process.}
    \label{fig:NN-t-FT}
\end{figure}

The spectral functions for each diagram in Fig.~\ref{fig:Feynman-All} are actually t-channel discontinuities cutting along the intermediate exchanged 2$\pi$, which is equivalent to s-channel cuts of the annihilation process $\bar{N}N\rightarrow \pi\pi \rightarrow \bar{N}N$, as is schematically shown in Fig.~\ref{fig:NN-t-FT}, taking the NLO football diagram as an example. It can be calculated via Cutkosky's cutting rules. However, the Cutkosky rules typically work in the s-channel c.m. frame. Therefore, to derive the TPE $NN$ amplitudes, we first calculate the spectral functions for the $\bar{N}N$ annihilation processes, then transform them to the forms required for the t-channel TPE $NN$ diagrams. Finally, we obtain the TPE $NN$ amplitudes by evaluating the dispersion integrals.  

The notations for the momentum for the $\bar{N}N$ processes are shown in Fig.~\ref{fig:NN-t-FT}
\begin{equation}\label{Notations1}
\begin{split}
   \tilde{p}_1=&\left(p_1^0,\mathbf{p}\right),~~~~ \tilde{p}_2=\left(p_2^0,-\mathbf{p}\right),\\  \tilde{p}_3=&\left(p_3^0,\mathbf{p}'\right),~~~~\tilde{p}_4=\left(p_4^0,-\mathbf{p}'\right), \\   \tilde{l}_1=&\left(l_1^0,\mathbf{l}\right),~~~~\tilde{l}_2=\left(l_2^0,-\mathbf{l}\right),\\ 
\end{split}
\end{equation}
with $\tilde{l}_1+\tilde{l}_2=\tilde{p}_1+\tilde{p}_2=\tilde{p}_3+\tilde{p}_4=\tilde{q}=(\mu,0)$. The components of each momentum above read
\begin{align}
   p_1^0&=p_2^0=\mu/2 ,~~|\mathbf{p}|=\sqrt{\mu^2/4-m^2},\nonumber\\
   p_3^0&=p_4^0=\mu/2 ,~~|\mathbf{p}'|=\sqrt{\mu^2/4-m^2},\nonumber\\
   l_1^0&=l_2^0=\mu/2 ,~~|\mathbf{l}|=\sqrt{\mu^2/4-\bf{m}_\pi^2}.
\end{align}

The amplitude for the $\bar{N}N$ football diagram in Fig.~\ref{fig:NN-t-FT}(b) reads
\begin{align}
   - i \mathcal{A}_{\bar{\text{N}}\text{N}}&=\frac{1}{2} \int \frac{d^4 l}{(2\pi)^4} \frac{\bar{v}_2 (\slashed{\tilde{l}}_1-\slashed{\tilde{l}}_2)u_1 \bar{u}_3 (\slashed{\tilde{l}}_1-\slashed{\tilde{l}}_2)v_4}{(\tilde{l}_1^2-m_\pi^2)(\tilde{l}_2^2-m_\pi^2)},
   \label{NNbarSDiract}
\end{align}
where $\tilde{l}_1=l$ and $\tilde{l}_2=\tilde{q}_3-l$.

With the Cutkosky rule, we have
\begin{align}
   - i \text{Disc}[\mathcal{A}_{\bar{\text{N}}\text{N}}]&=2\text{Im}[\mathcal{A}_{\bar{\text{N}}\text{N}}] \nonumber\\
   &= \text{Disc}\left[ \frac{1}{2}\int \frac{d^4 l}{(2\pi)^4} \frac{V_{\mathrm{L}}\cdot V_{\mathrm{R}}}{(\tilde{l}_1^2-m_\pi^2)(\tilde{l}_2^2-m_\pi^2)}\right]\nonumber \\
   &=\frac{1}{2} \int \frac{d^4 l}{(2\pi)^4} V_{\mathrm{L}}\cdot V_{\mathrm{R}} (-2\pi i)\delta(\tilde{l}_1^2-m_\pi^2)\nonumber \\
   &\cdot(-2\pi i)\delta(\tilde{l}_2^2-m_\pi^2)\nonumber\\
   &=\frac{1}{2}\int d\Omega \frac{-|\mathbf{l}|}{16\pi^2\mu} V_{\mathrm{L}}\cdot V_{\mathrm{R}} ,
   \label{eq:NNbarCUT}
\end{align}
where we have used $V_{\mathrm{L}}\cdot V_{\mathrm{R}}$ to denote the numerator for simplification. 
\begin{align}
   V_{\mathrm{L}}= &\bar{v}_2 (2\slashed{l}-\slashed{\tilde{q}})u_1, \nonumber\\
   V_{\mathrm{R}}= &\bar{u}_3 (2\slashed{l}-\slashed{\tilde{q}})v_4, 
\end{align}
\begin{align}
   V_{\mathrm{L}}\cdot V_{\mathrm{R}} &= 4\bar{v}_2 \slashed{l} u_1 \bar{u}_3\slashed{l}v_4 + 2\bar{v}_2 \slashed{l} u_1 \bar{u}_3(-\slashed{\tilde{q}})v_4 \nonumber\\
   &+ 2\bar{v}_2 (-\slashed{\tilde{q}}) u_1 \bar{u}_3\slashed{l}v_4 +\bar{v}_2 (-\slashed{\tilde{q}}) u_1 \bar{u}_3(-\slashed{\tilde{q}})v_4.
\end{align}

In the expression above, we only need to deal with three different integrals
\begin{align}
   T_0=&\int \frac{d^4 l}{(2\pi)^4} \frac{1}{(l^2-m_\pi^2)((l-\tilde{q})^2-m_\pi^2)},\nonumber\\
   T_1^\mu=&\int \frac{d^4 l}{(2\pi)^4} \frac{l^\mu} {(l^2-m_\pi^2)((l-\tilde{q})^2-m_\pi^2)},\nonumber\\
   T_2^{\mu\nu}=&\int \frac{d^4 l}{(2\pi)^4} \frac{l^\mu l^\nu}{(l^2-m_\pi^2)((l-\tilde{q})^2-m_\pi^2)}.
\end{align}
Following Eq.~(\ref{eq:NNbarCUT}), the discontinuity for each integral reads
\begin{align}
   \text{Disc}[T_0]=&\int d\Omega \frac{-|\mathbf{l}|}{16\pi^2\mu} \nonumber\\
   =&-\frac{|\mathbf{l}|}{4\pi \mu}\equiv B_0,\\
   \text{Disc}[T_1^\mu]=&\tilde{q}^\mu \int d\Omega \frac{-|\mathbf{l}|}{16\pi^2\mu}\frac{l\cdot \tilde{q}}{\tilde{q}^2} \nonumber\\
   =&  -\tilde{q}^\mu\frac{|\mathbf{l}|}{8\pi \mu} \equiv \tilde{q}^\mu B_1,\\
   \text{Disc}[T_2^{\mu\nu}]=&g^{\mu\nu} \int d\Omega \frac{-|\mathbf{l}|}{16\pi^2(D-1)\mu}\frac{l^2 \tilde{q}^2-l\cdot \tilde{q}^2}{\tilde{q}^2} \nonumber\\
   +& \tilde{q}^\mu \tilde{q}^\nu \int d\Omega \frac{-|\mathbf{l}|}{16\pi^2(D-1)\mu}\frac{D l\cdot \tilde{q}^2-l^2 \tilde{q}^2}{(\tilde{q}^2)^2} \nonumber\\
   =&g^{\mu\nu} \frac{|\mathbf{l}|^3}{12\pi \mu} - \tilde{q}^\mu \tilde{q}^\nu \frac{3\mu^2|\mathbf{l}|+4 |\mathbf{l}|^3}{48\pi \mu^3} \nonumber\\
   \equiv & g^{\mu\nu} B_{21}+\tilde{q}^\mu \tilde{q}^\nu B_{22},
\end{align}
where we have taken $D=4$, the dimensionality of the Feynman integrals.

With the redefinition above, we can rewrite the loop integral in Eq.~(\ref{eq:NNbarCUT}) as
\begin{align}
   &\text{Im}[\mathcal{A}_{\bar{\text{N}}\text{N}}] \nonumber\\
   =&B_{21}\bar{v}_2 \gamma^\mu u_1 \bar{u}_3 \gamma^\mu v_4 +  B_{22}\bar{v}_2 \slashed{\tilde{q}} u_1 \bar{u}_3 \slashed{\tilde{q}} v_4 \nonumber\\
   -&  B_1\bar{v}_2 \slashed{\tilde{q}} u_1 \bar{u}_3\slashed{\tilde{q}}v_4 +\frac{1}{4}B_0 \bar{v}_2 \slashed{\tilde{q}} u_1 \bar{u}_3\slashed{\tilde{q}}v_4.
\end{align}

Next, we transform these spectral functions of $\bar{N}N$ into those of $NN$ with the following replacement,
\begin{align}
   u_1(\tilde{p}_1)\rightarrow u_1(p_1),~~ &\bar{v}_2(\tilde{p}_2)\rightarrow \bar{u}_3(p_3), \nonumber\\
   \bar{u}_3(\tilde{p}_3)\rightarrow \bar{u}_4(p_4),~~&v_4(\tilde{p}_4)\rightarrow u_2(p_2), \nonumber\\
   \tilde{p}_1 \rightarrow p_1,~~ 
   \tilde{p}_2 \rightarrow -p_3,~~&
   \tilde{p}_3 \rightarrow -p_2,~~ 
   \tilde{p}_4 \rightarrow p_4, \nonumber\\
   \tilde{q}=\tilde{p}_1+\tilde{p}_2 \rightarrow & q=p_1-p_3.
\end{align}
Note that this transformation incorporates both the nucleon-antinucleon replacement ($v\rightarrow u$) and the frame conversion from the two-pion c.m. frame to the two-nucleon c.m. frame. The $B_0$, $B_1$, $B_{21}$, and $B_{22}$ remain invariant in this transformation.

Now, we obtain the spectral functions as
\begin{align}
   &\text{Im}[\mathcal{A}_{\text{NN}}] \nonumber\\  
   =&B_{21}\bar{u}_3 \gamma^\mu u_1 \bar{u}_4 \gamma^\mu u_2  + \frac{B_0-4 B_1+4 B_{22}}{4}\bar{u}_3 \slashed{q} u_1 \bar{u}_4 \slashed{q} u_2  .
\end{align}
Then, $\mathcal{A}_{\text{NN}}$ can be obtained straightforwardly via dispersion integrals
\begin{align}\label{eq:SFRDspint}
   \mathcal{A}_{\text{NN}}(t) &=A(-|\mathbf{q}|^2) \nonumber\\
   &=P(t)+\frac{1}{\pi}\int_{4m_\pi^2} \text{d}\mu^2 \left(\frac{-|\mathbf{q}|^2}{\mu^2}\right)^n \frac{\text{Im}[\mathcal{A}_{\text{NN}}(\mu^2)]}{\mu^2+|\mathbf{q}|^2}    
\end{align}
where $t=(p_1-p_3)^2=-|\mathbf{q}|^2$ is the Mandelstam symbol for $NN$ scattering, $P(t)$ is the subtracted polynomial, and $n$ denotes the subtraction order. 

Similarly, the general expressions take the following form
\begin{align}
   \mathcal{A}_{\text{NN}}(t) &= \sum B_i L_i,
\end{align}
where $B_i$ denotes different combinations of bilinear and $L_i$ denotes the dispersion integrals for each Feynman diagram of the form shown in Eq.~(\ref{eq:SFRDspint}). The isospin and coupling factors are included in the $L_i$s. For each type of TPE diagrams in Fig.~\ref{fig:Feynman-All}, the related spectral functions can be obtained starting from the corresponding $\bar{N}N$ amplitudes as
\begin{align}
   \mathcal{A}_{\bar{\text{N}}\text{N}}^{\text{FT}} &=  \frac{i}{2}\int \frac{d^4 l}{(2\pi)^4} \frac{V_{\mathrm{L}}^{\mathrm{FT}}\cdot V_{\mathrm{R}}^{\mathrm{FT}}}{(\tilde{l}_1^2-m_\pi^2)(\tilde{l}_2^2-m_\pi^2)},\nonumber \\
   \mathcal{A}_{\bar{\text{N}}\text{N}}^{\text{TRI}} &= i \int \frac{d^4 l}{(2\pi)^4} \frac{V_{\mathrm{L}}^{\mathrm{TRI}}\cdot V_{\mathrm{R}}^{\mathrm{TRI}}}{(\tilde{l}_1^2-m_\pi^2)(\tilde{l}_2^2-m_\pi^2)((\tilde{l}_1-\tilde{p}_1)^2-m^2)},\nonumber \\
   \mathcal{A}_{\bar{\text{N}}\text{N}}^{\text{TRII}} &=i \int \frac{d^4 l}{(2\pi)^4} \frac{V_{\mathrm{L}}^{\mathrm{TRII}}\cdot V_{\mathrm{R}}^{\mathrm{TRII}}}{(\tilde{l}_1^2-m_\pi^2)(\tilde{l}_2^2-m_\pi^2)((\tilde{l}_1-\tilde{p}_3)^2-m^2)},\nonumber \\
   \mathcal{A}_{\bar{\text{N}}\text{N}}^{\text{BOX}} &=i \int \frac{d^4 l}{(2\pi)^4} \frac{V_{\mathrm{L}}^{\mathrm{BOX}}\cdot V_{\mathrm{R}}^{\mathrm{BOX}}}{(\tilde{l}_1^2-m_\pi^2)(\tilde{l}_2^2-m_\pi^2)}\nonumber \\
   &\cdot\frac{1}{((\tilde{l}_1-\tilde{p}_1)^2-m^2)((\tilde{l}_1-\tilde{p}_3)^2-m^2)} ,\nonumber\\
   \mathcal{A}_{\bar{\text{N}}\text{N}}^{\text{CRO}} &=i \int \frac{d^4 l}{(2\pi)^4} \frac{V_{\mathrm{L}}^{\mathrm{CRO}}\cdot V_{\mathrm{R}}^{\mathrm{CRO}}}{(\tilde{l}_1^2-m_\pi^2)(\tilde{l}_2^2-m_\pi^2)}\nonumber \\
   &\cdot\frac{1}{((\tilde{l}_1-\tilde{p}_1)^2-m^2)((\tilde{l}_1-\tilde{p}_4)^2-m^2)} .
\end{align}

\begin{align}
   V_{\mathrm{R}}^{\mathrm{FT}}&= V_{\mathrm{R}}^{\mathrm{TRI}}=i\bar{u}_3(\slashed{l}-\slashed{\tilde{p}}_1-\slashed{\tilde{p}}_2)v_4, \nonumber\\
   V_{\mathrm{R}}^{\mathrm{TRII}}&=V_{\mathrm{R}}^{\mathrm{BOX}} \nonumber\\
   &=i\bar{u}_3\gamma^5(\slashed{l}-\slashed{\tilde{p}}_3 -\slashed{\tilde{p}}_4)(\slashed{l}-\slashed{\tilde{p}}_3+m)\gamma^5\slashed{l}v_4 ,\nonumber\\
   V_{\mathrm{R}}^{\mathrm{CRO}}&=i\bar{u}_3\gamma^5\slashed{l}(\slashed{\tilde{p}}_4-\slashed{l}+m)\gamma^5(\slashed{l}-\slashed{\tilde{p}}_3 -\slashed{\tilde{p}}_4)v_4.
\end{align}
All the $V_{\mathrm{L}}$ above can be written in the most general form as follows
\begin{align}
   V_{\mathrm{L}} &= \mathcal{T}_1\bar{v}_2u_1+\mathcal{T}_2\bar{v}_2\slashed{l}u_1+\mathcal{T}_3\bar{v}_2\slashed{\tilde{p}}_1u_1+\mathcal{T}_4\bar{v}_2\slashed{l}\slashed{\tilde{p}}_1u_1 \nonumber\\
   &+\mathcal{T}_5\bar{v}_2\slashed{\tilde{p}}_2u_1 +\mathcal{T}_6\bar{v}_2\slashed{\tilde{p}}_2\slashed{l}u_1 +\mathcal{T}_7\bar{v}_2\slashed{\tilde{p}}_2\slashed{\tilde{p}}_1u_1 +\mathcal{T}_8\bar{v}_2\slashed{\tilde{p}}_2\slashed{l}\slashed{\tilde{p}}_1u_1,
\end{align}
where $\mathcal{T}_{1\dots8}$ are the corresponding components of $\bar{N}N\rightarrow\pi\pi$ amplitudes for each diagram. Notably, in the above derivation, we have explicitly demonstrated the Lorentz structures of the momenta in $V_{\mathrm{L}}$ and $V_{\mathrm{R}}$ without further simplification via the equation of motion. This is indispensable for the non-perturbative treatments in future work. The expressions for $\mathcal{T}_{1\dots8}$ are too complex to present explicitly in the manuscript~\footnote{The expressions can be provided as a Mathematica notebook from the authors upon reasonable request.}.

Several points need further attention. First, at N$^3$LO, the relevant N$^2$LO $\pi N$ amplitudes denoted by the shadowed circles or blocks in Fig.~\ref{fig:N3LO-part1} and Fig.~\ref{fig:N3LO-part2} should be properly renormalized. In addition to the wave function renormalization, which is already exhibited explicitly in Fig.~\ref{fig:N3LO-part1} and Fig.~\ref{fig:N3LO-part2}, the mass of the propagating nucleon and the $\pi NN$ vertices of the LO Born terms should also be renormalized up to N$^2$LO. As a consequence, we adopt $g_A=1.267$ for the renormalized $\pi N$ amplitudes. Additionally, the chiral corrections to the decay constant should be taken into account. Details regarding the renormalization of the $\pi N$ amplitudes can be found in Ref.~\cite{Chen:2012nx}, while those for SU(3) cases are available in Ref.~\cite{Lu:2018zof}.

Owing to the induced form factor specified in Eq.~(\ref{eq:FormFactor}), the dispersion integrals of Eq.~(\ref{eq:SFRDspint}) are free of divergences. To ensure that the corresponding contact terms can renormalize the subtracted polynomials, we select the subtraction order $n$ in Eq.~(\ref{eq:SFRDspint}) as described below. The partial-wave projection will result in the bilinear outcome being zero after multiplication by a polynomial of $q^2$ of a certain order. Taking $\mathcal{B}=\bar{u}_3\gamma^\mu u_1~\bar{u}_4\gamma_\mu u_2$ as an example, after projection onto partial waves with total angular momentum $J=3$, $\mathcal{B}\cdot(a+b q^2)$ vanishes, which means that one actually obtains the same results for subtraction orders for $n=0,1,2$. Similarly, one can obtain the same results for $n=0,1,2,3$ when projected to partial waves with $J=4$. We then traverse all bilinear structures and take the maximum achievable order $n$, i.e., $n^{(3)}_{\mathrm{max}}$ for the $J=3$ partial wave as the subtraction order of the dispersion integral corresponding to each bilinear structure. For the $J\le2$ partial waves, all TPE contributions arise starting from NLO, which contain the $\mathcal{O}(p^2)$ contact terms. Thus, we are safe to use any subtraction order $n$ satisfying $0\le n<n^{(3)}_{\mathrm{max}}$.

\section{Perturbative $NN$ scattering in peripheral partial waves}\label{sec:results}

In the present manuscript, we concentrate on the peripheral partial waves with total angular momentum $3\le J\le 5$. These partial waves probe the long- and intermediate-range components of the nuclear force and exhibit no sensitivity to short-range interactions, even though at N$^3$LO, the $J=3$ partial waves already incorporate short-range contact terms according to the power counting rules~\cite{Xiao:2018jot}. Additionally, non-perturbative effects in these partial waves are not relevant, and a perturbative treatment leads to the phase shifts and the mixing angle for each partial wave as follows~\cite {Xiao:2020ozd},
\begin{align}
   \delta_{LSJ}=&-\frac{m^2|\mathbf{p}|}{16\pi^2 E}\text{Re}\langle LSJ|\mathcal{A}_{\mathrm{NN}}|LSJ\rangle \nonumber\\
   \epsilon_J=&\frac{m^2|\mathbf{p}|}{16\pi^2 E}\text{Re}\langle J-1,1,J|\mathcal{A}_{\mathrm{NN}}|J+1,1,J\rangle
\end{align}
where $|\mathbf{p}|$ is the c.m. momentum of $NN$ scattering and $E=\sqrt{m^2+|\mathbf{p}|^2}$ is the energy of the incoming or outgoing nucleon.  $L$ is the orbital angular momentum, and $S$ is the total spin.

The results are free of any parameters except the momentum cutoff for the dispersion integrals. We adopt a common cutoff for the dispersion integrals from all chiral orders and vary it from $0.5$ GeV to $0.8$ GeV, which covers the typical choices in the construction of chiral nuclear forces. We have verified that even with a cutoff of 1.5 GeV, the results remain almost unchanged, except for a slightly broader uncertainty band.

\subsection{$J=3$ partial waves}

\begin{figure*}[htbp]
\centering
\includegraphics[width=0.35\textwidth]{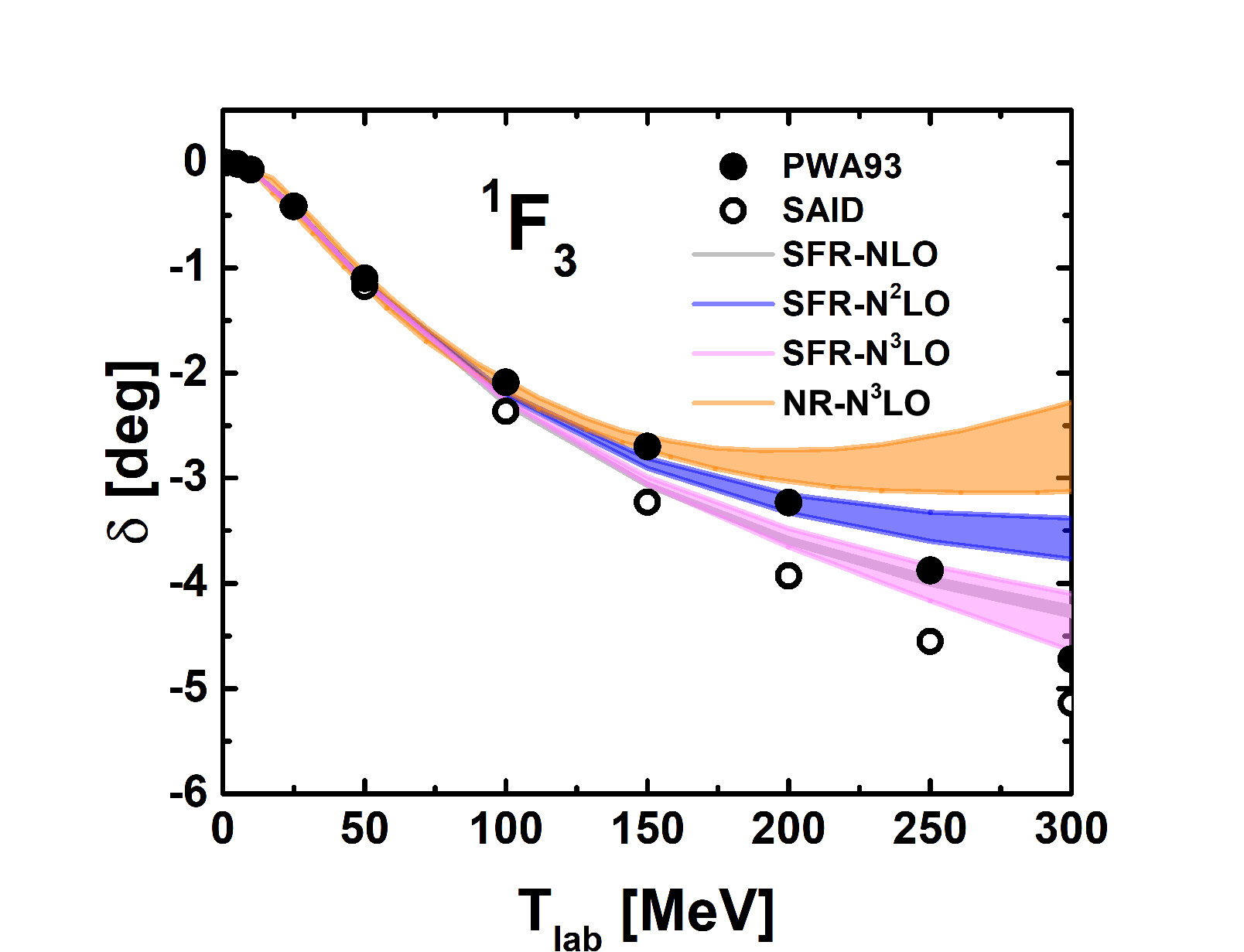}\hspace{-8mm}
\includegraphics[width=0.35\textwidth]{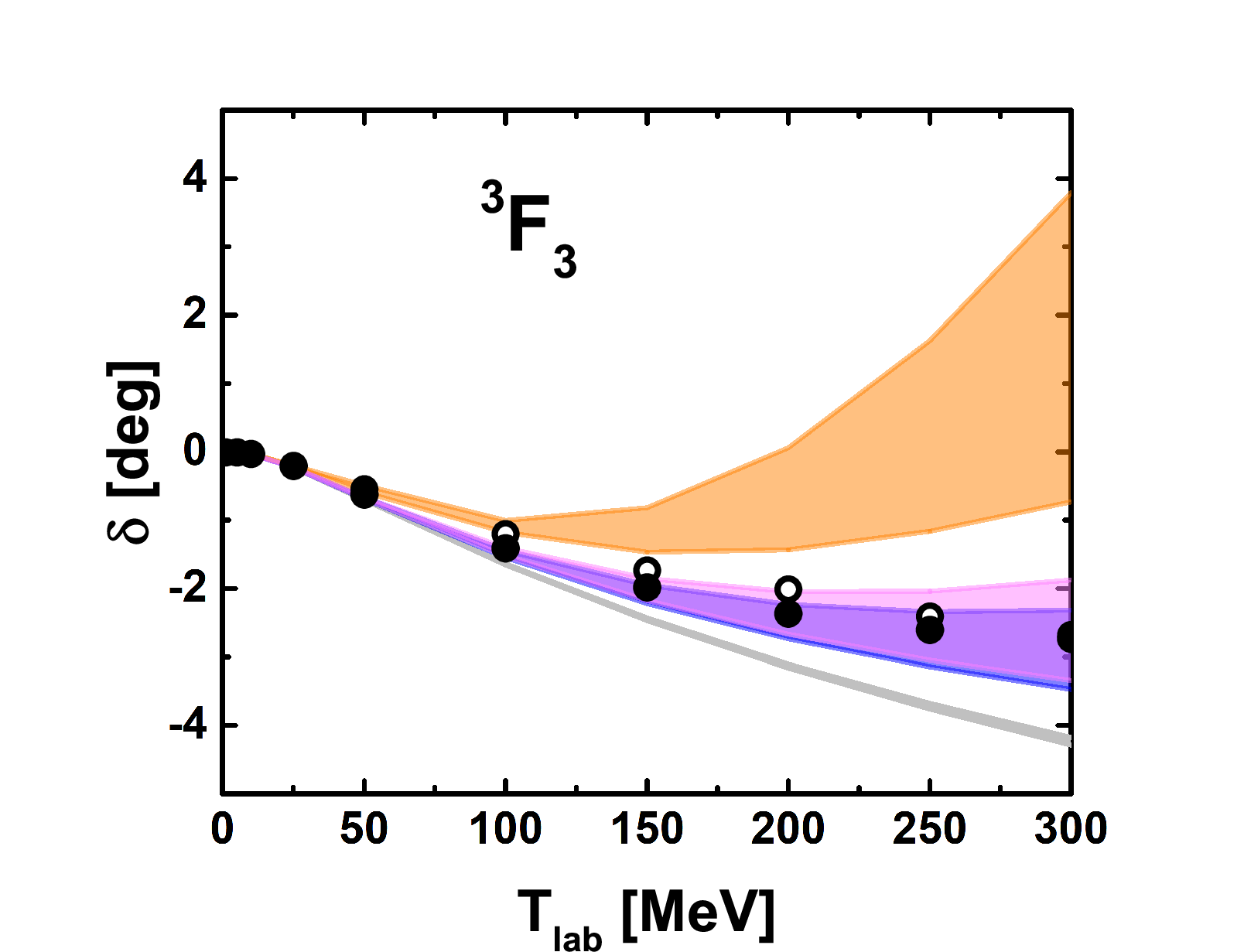}\hspace{-8mm}
\includegraphics[width=0.35\textwidth]{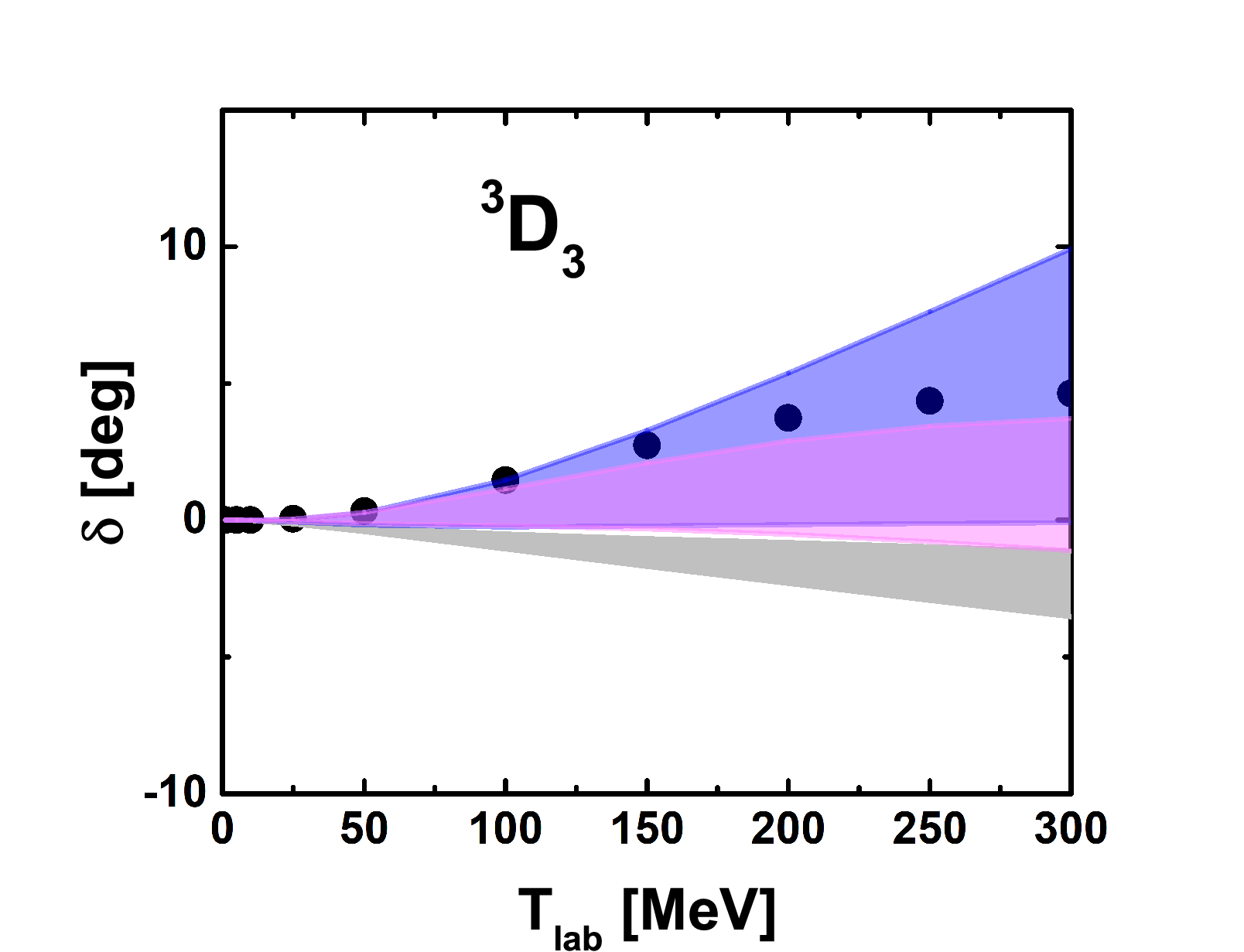}\\ 
\includegraphics[width=0.35\textwidth]{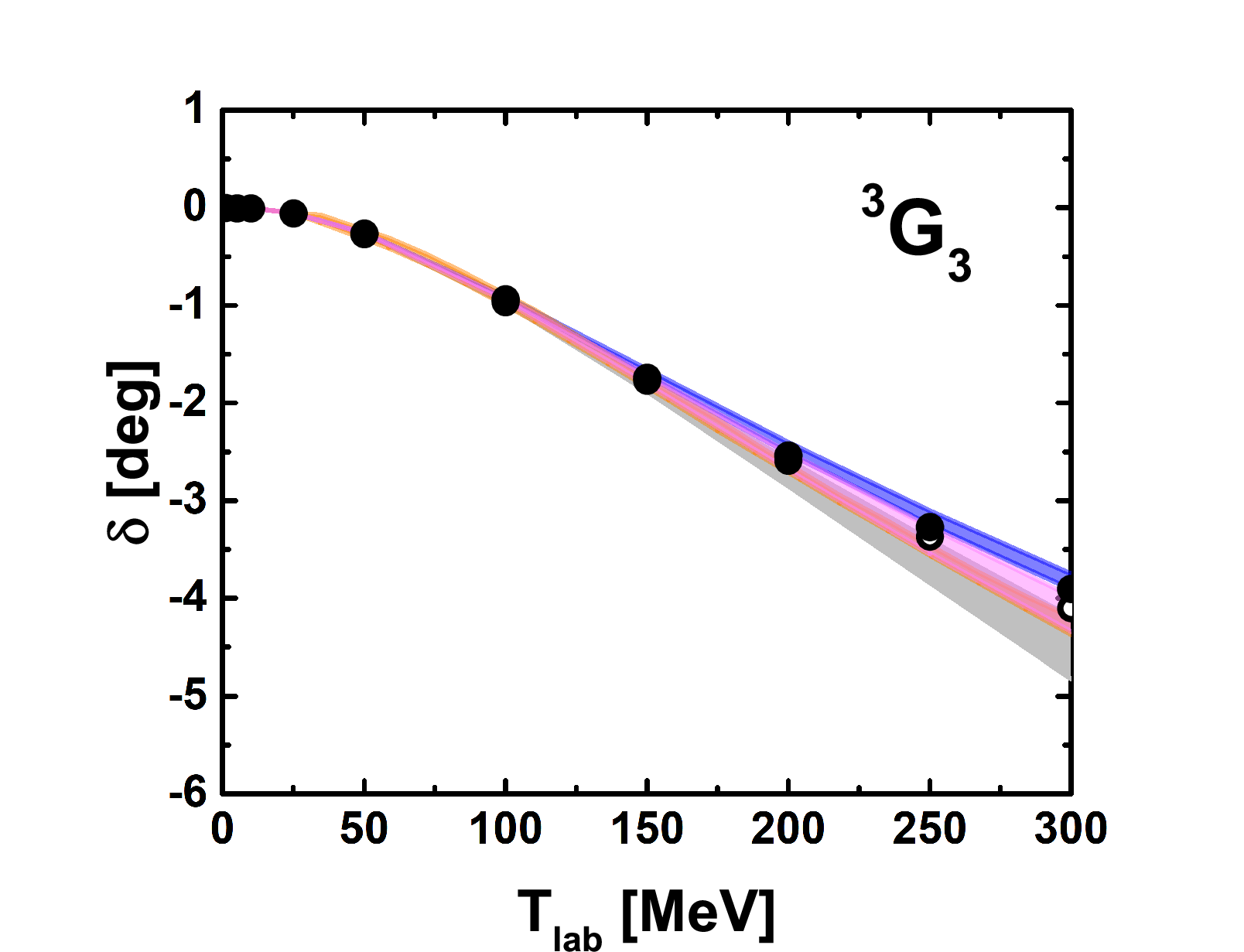}\hspace{-8mm}
\includegraphics[width=0.35\textwidth]{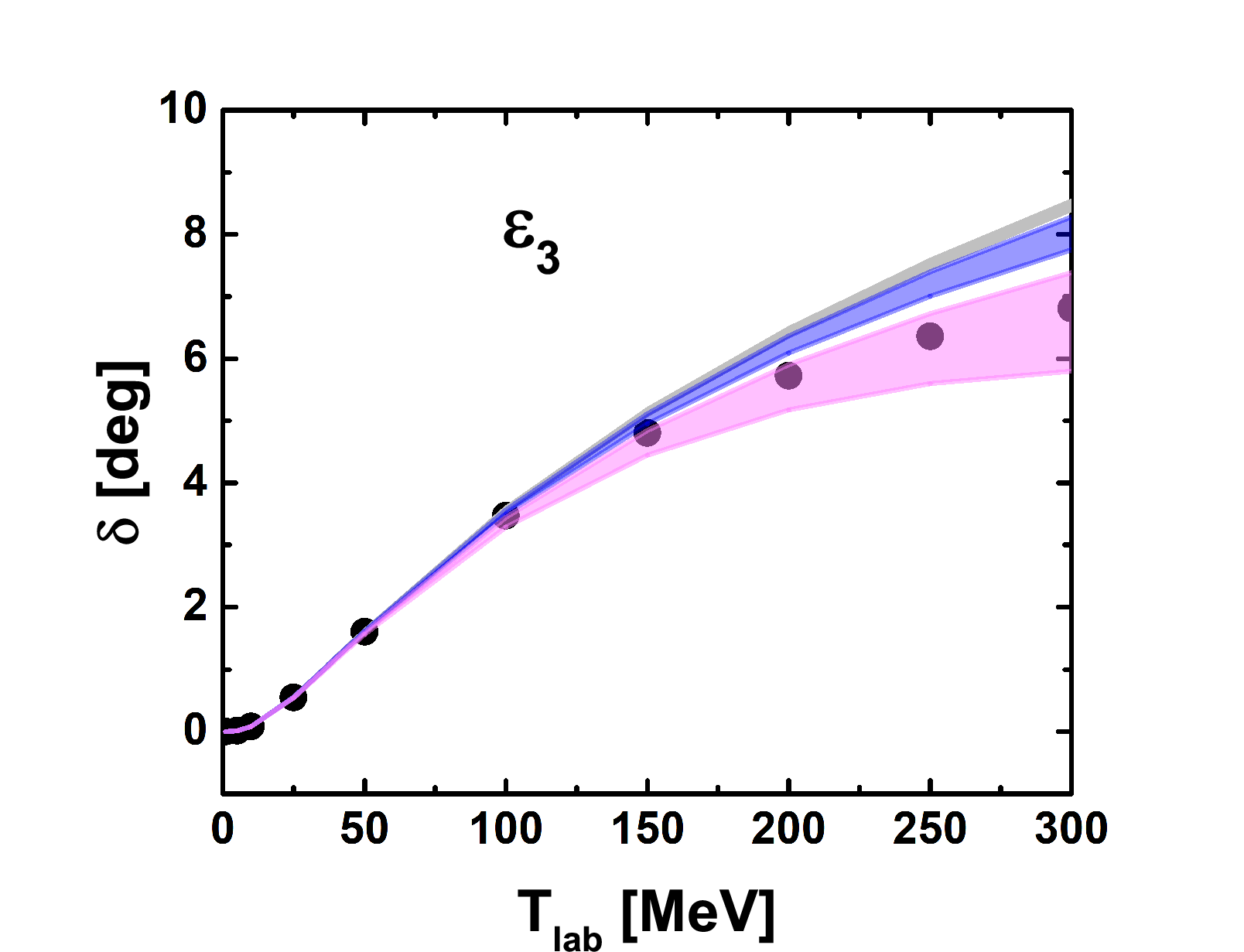}
\caption{Phase shifts and mixing angle for the $J=3$ partial waves. The gray, blue, and magenta bands are results from relativistic TPE $NN$ interactions at NLO, N$^2$LO, and N$^3$LO, respectively, with a cutoff in the range $[0.5, 0.8]$ GeV. For comparison, the N$^3$LO non-relativistic results are shown with orange bands (the two $F$-waves from Ref.~\cite{Entem:2014msa} and the $G$-waves from Ref.~\cite{Entem:2015xwa}). The solid and open dots represent the data from the Nijmegen multi-energy neutron-proton (n-p) phase shift analysis~\cite{Stoks:1993tb} and the VPI/GWU single-energy n-p analysis SM99~\cite{Arndt:1994br}, respectively. }
\label{fig:PSJ3}
\end{figure*}

The phase shifts for the $J=3$ partial waves and the mixing angle $\epsilon_3$ at NLO, N$^2$LO, and N$^3$LO are shown in Fig.~\ref{fig:PSJ3}. Note that in these figures we present the partial-wave analyses from the Nijmegen~\cite{Stoks:1993tb} and SAID~\cite{Arndt:1994br} groups. In the lower-energy region, e.g., $T_{\mathrm{lab}}<100$ MeV, both analyses yield nearly identical phase shifts, whereas in the higher-energy region, significant discrepancies arise in some partial waves, especially in $^1F_3$. These discrepancies can be regarded as uncertainties in the phase shifts to some extent. However, since the phase shifts are not true observables, a meaningful comparison ought to be performed using the $\chi^2/\mathrm{d.o.f.}$ for $NN$ scattering observables. We leave this to future work once the LECs are pinned down for the partial waves with $J\le2$.

Within the relativistic framework, we find good agreement with both the PWA93 and SAID data for the $^1F_3$, $^3G_3$, and $\epsilon_3$ partial waves, even at NLO.  However, at higher energies, incorporating higher-order contributions results in a notable improvement, especially at N$^3$LO in the $^3F_3$ and $\epsilon_3$ partial waves. In the $^1F_3$ partial wave, the contribution from N$^2$LO appears to be canceled by those from N$^3$LO, and up to N$^3$LO, the results overlap almost with the NLO ones. On the other hand, the additional corrections arising from extending the calculation from N$^2$LO to N$^3$LO are comparatively small in general. We can see that N$^2$LO and N$^3$LO phase shifts for the $^3F_3$ and $^3G_3$ partial waves almost overlap. 

In the $J=3$ partial waves, the $^3D_3$ partial wave is an exception. It is widely recognized that the short-range component, as well as the non-perturbative effect, is crucial for this partial wave~\cite{Epelbaum:1999dj}. In the non-relativistic framework, LECs contributing to $D$-waves only emerge at N$^3$LO, and their absence makes the results sensitive to the cutoff. Consequently, such a perturbative treatment incorporating only the long- and intermediate-range components for the $^3D_3$ partial wave has limited reference value. We show it here just for completeness.

In Fig.~\ref{fig:PSJ3}, we also present the non-relativistic results up to N$^3$LO for the $^1F_3$, $^3F_3$~\cite{Entem:2014msa}, and $^3G_3$~\cite{Entem:2015xwa} partial waves, which are obtained from the Idaho group and represented by orange bands. It should be noted that the non-relativistic results for $^1F_3$ and $^3F_3$ correspond to a cutoff variation in the range $[0.7,1.5]$ GeV~\cite{Entem:2014msa}, whereas those for $^3G_3$ correspond to a cutoff range of $[0.7,0.9]$ GeV~\cite{Entem:2015xwa}. We note that sharp cutoffs are used in both works, so one needs to be careful when comparing them directly. Note that the non-relativistic results for $^3D_3$ and $\epsilon_3$ are absent, owing to the reason explained above.

\subsection{$J=4$ partial waves}

\begin{figure*}[htbp]
\centering
\includegraphics[width=0.35\textwidth]{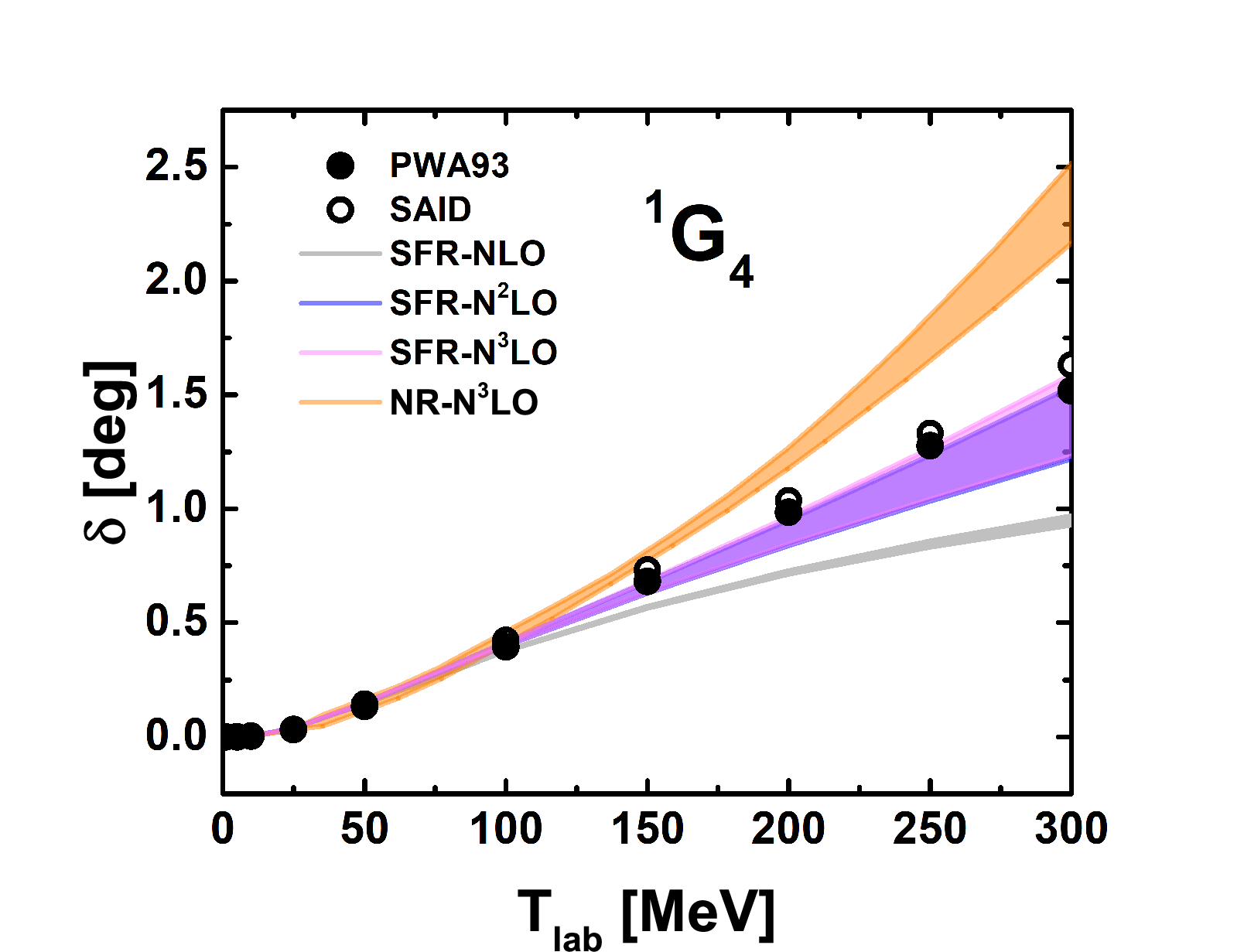}\hspace{-8mm}
\includegraphics[width=0.35\textwidth]{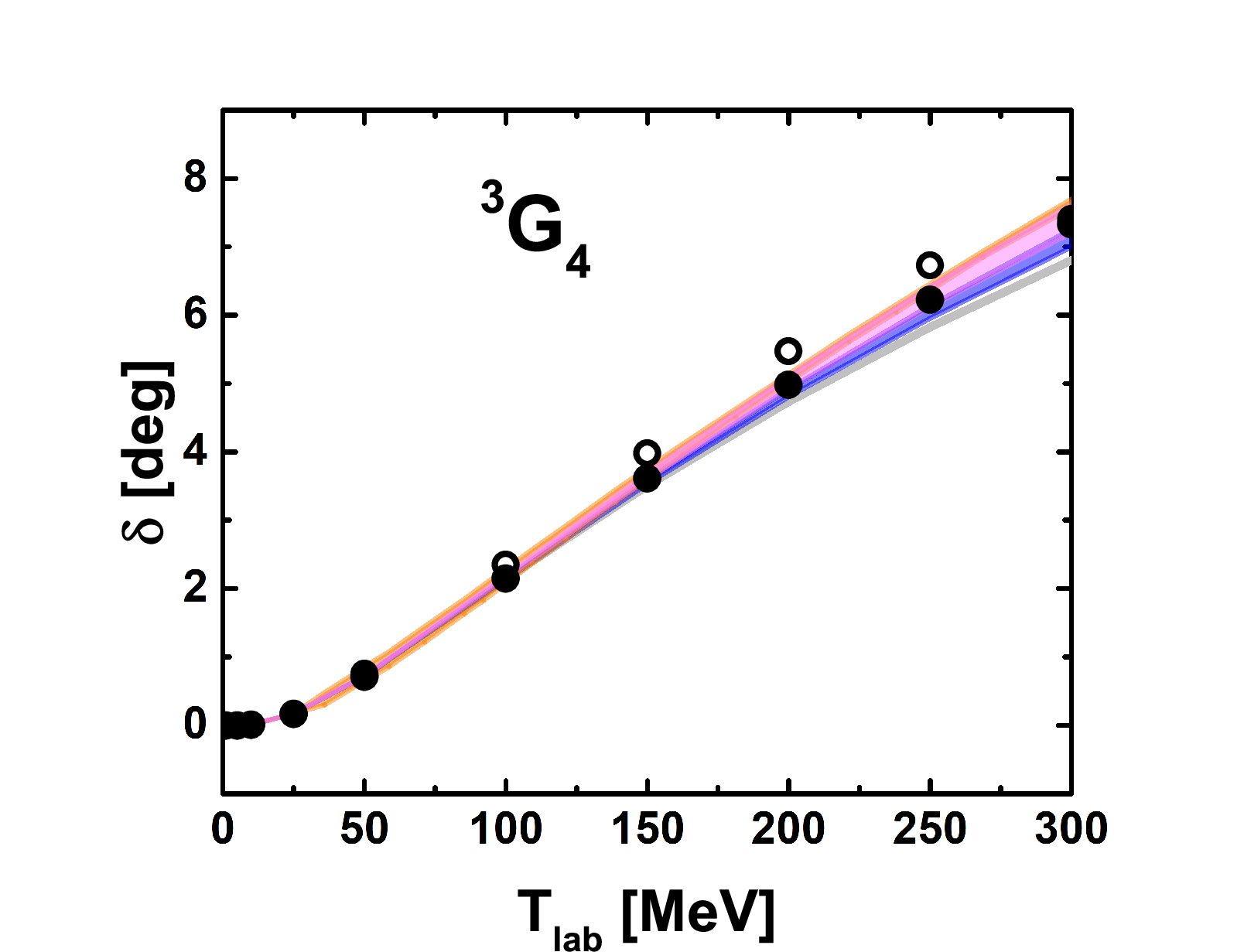}\hspace{-8mm}
\includegraphics[width=0.35\textwidth]{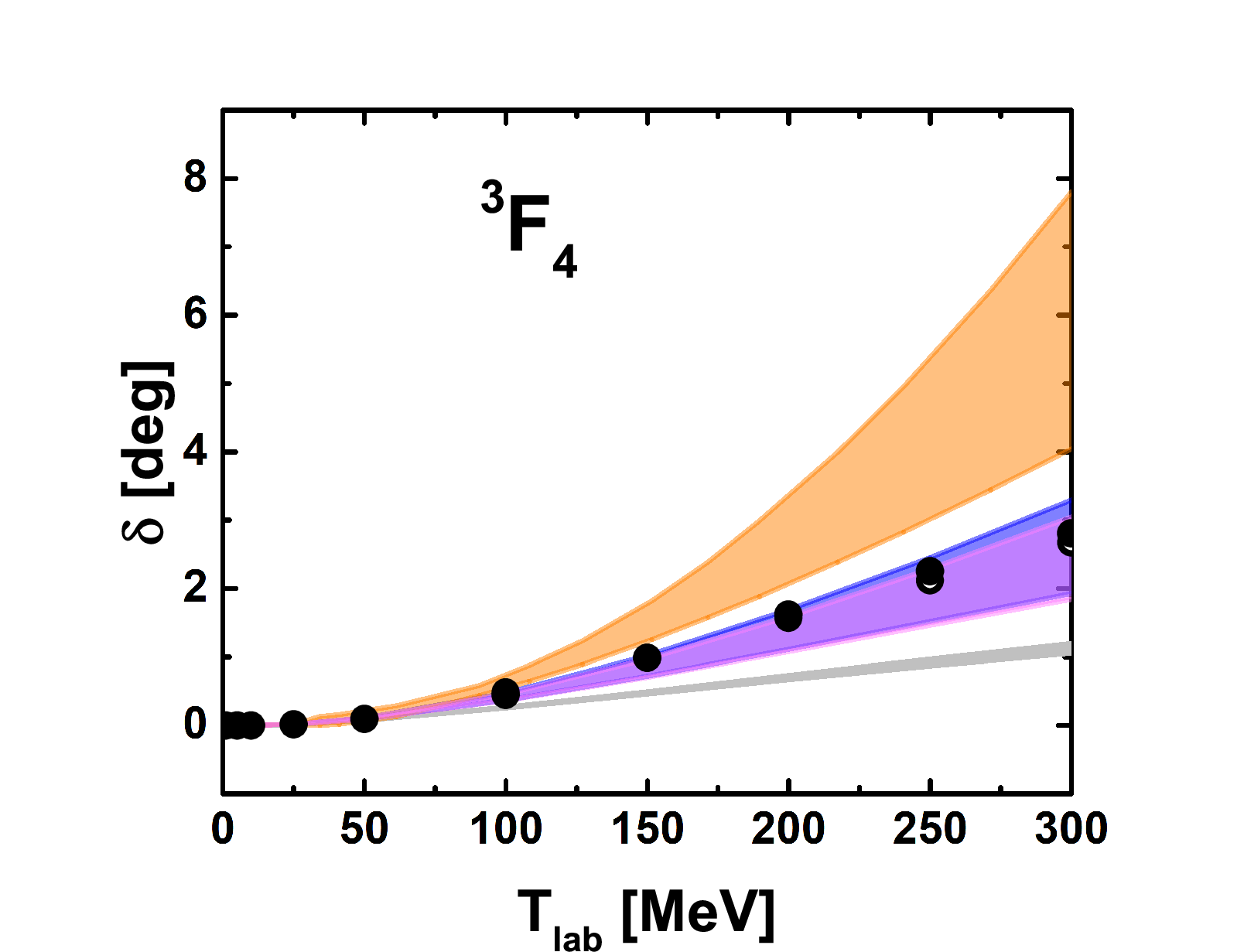}\\ 
\includegraphics[width=0.35\textwidth]{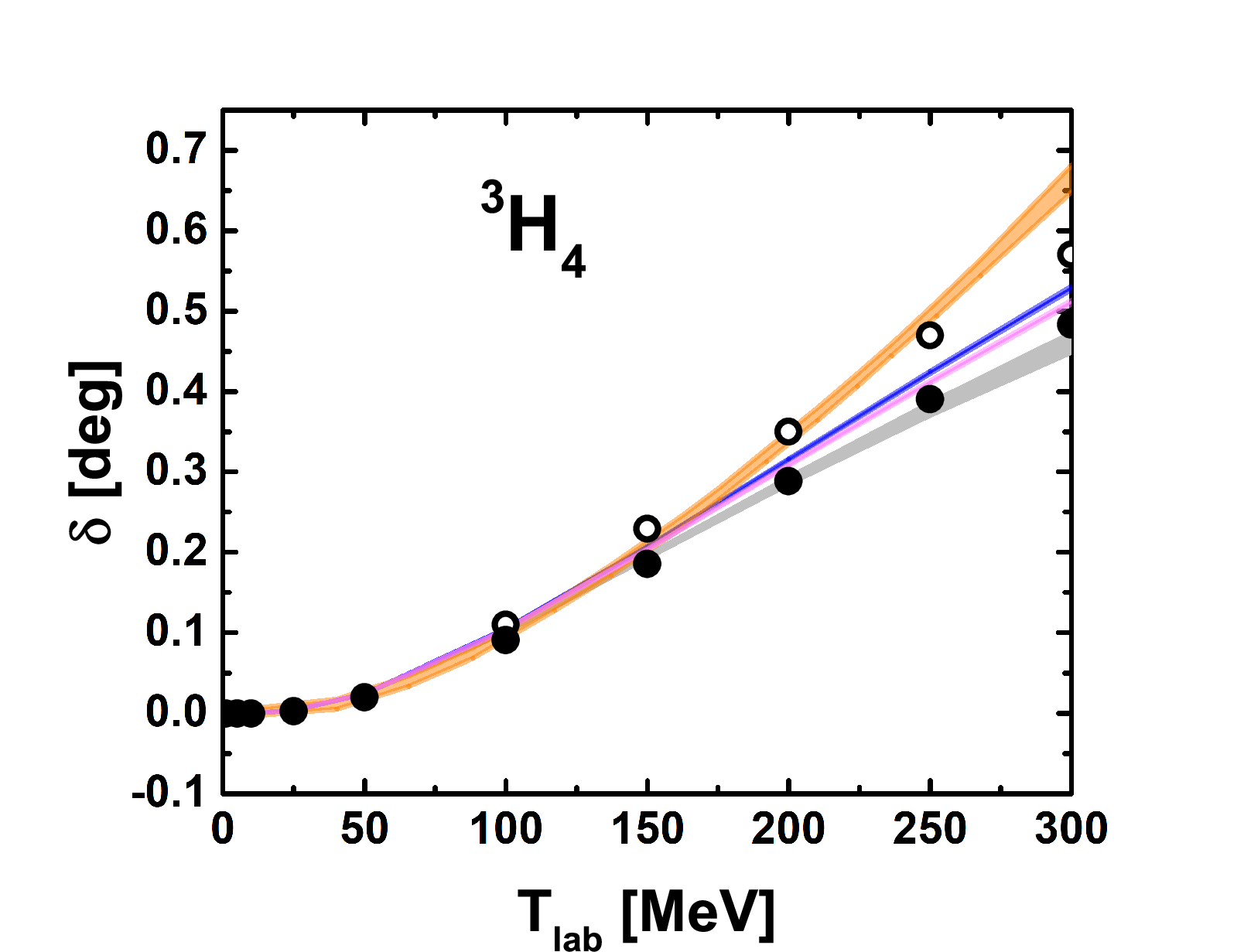}\hspace{-8mm}
\includegraphics[width=0.35\textwidth]{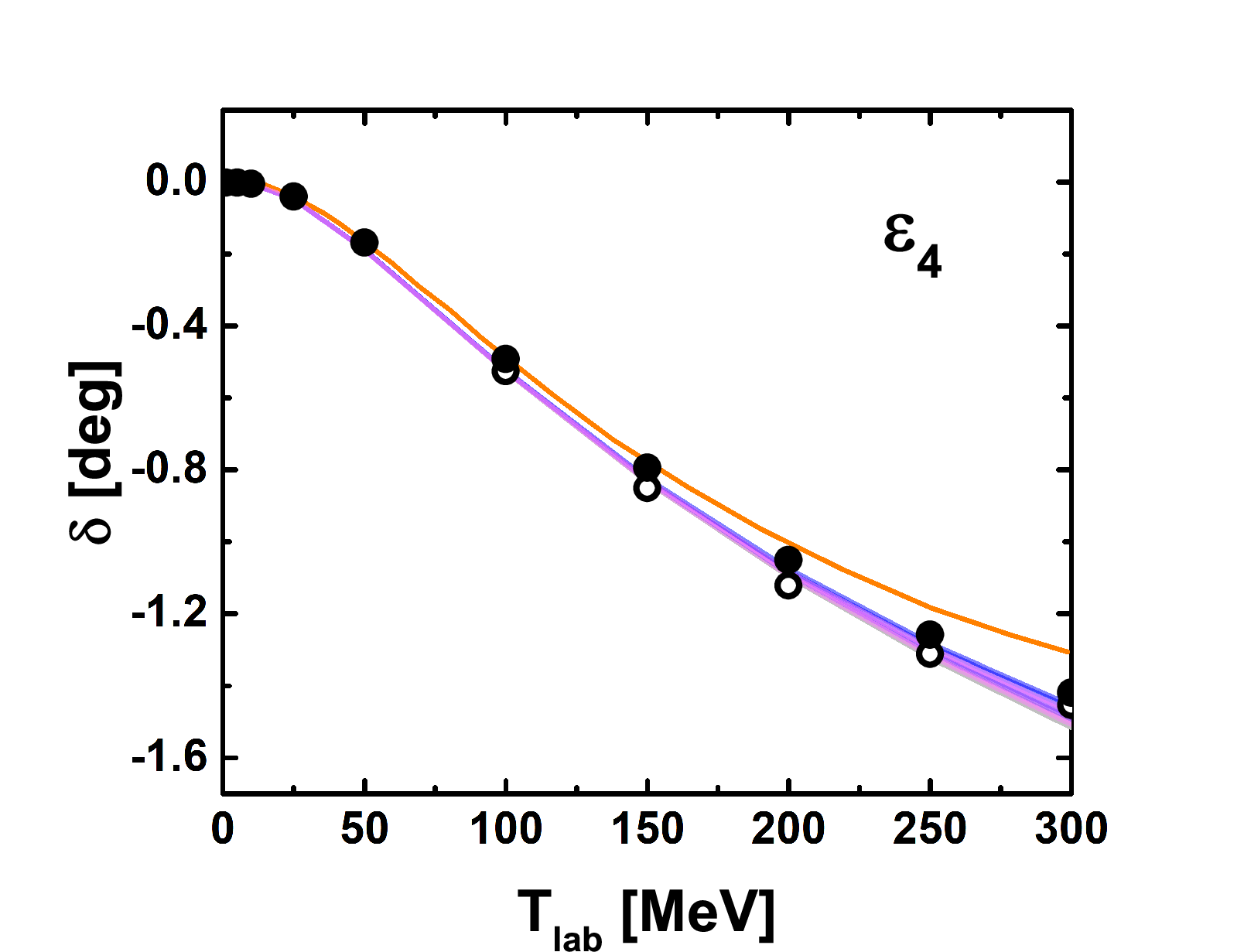}\\ 
\caption{Same as Fig.~\ref{fig:PSJ3} but for the $J=4$ partial waves. The solid and open dots are the data from the Nijmegen multi-energy n-p phase shift analysis~\cite{Stoks:1993tb} and the GWU n-p analysis SP07~\cite{Arndt:2007qn}. }
\label{fig:PSJ4}
\end{figure*}

The phase shifts for the $J=4$ partial waves and the mixing angle $\epsilon_4$ are shown in Fig.~\ref{fig:PSJ4}. For these partial waves, we find a similar conclusion to that in the $J=3$ partial waves. Up to NLO, the relativistic results are in good agreement with the PWA93 or SAID data only in the lower-energy region, and significant improvements are observed at higher energies when the calculation is extended to higher orders. Besides, we find that in all these $J=4$ partial waves, calculations up to N$^2$LO and N$^3$LO yield almost the same phase shifts. Such a pattern actually reflects good convergence of the relativistic chiral nuclear force, indicating that for these partial waves at least, chiral nuclear forces truncated at N$^2$LO are already sufficient. 

In Fig.~\ref{fig:PSJ4}, we also show the non-relativistic results up to N$^3$LO. Note that the non-relativistic results for the $^3F_4$ partial wave are from Ref.~\cite{Entem:2014msa} and the others are from Ref.~\cite{Entem:2015xwa}. Besides, the non-relativistic mixing angle $\epsilon_4$ result was obtained with a cutoff $\Lambda=0.8$ GeV~\cite{Entem:2015xwa}. In non-relativistic calculations, the N$^3$LO results can still achieve good agreement with the PWA93 or SAID data only in the lower-energy region, e.g., in $^1G_4$ and $^3F_4$. In these partial waves, the description for the higher energy region is significantly improved in the next order, i.e., N$^4$LO~\cite{Entem:2014msa,Entem:2015xwa}. Even the corrections at N$^5$LO can barely be neglected~\cite{Entem:2015xwa}.

\subsection{$J=5$ partial waves}

\begin{figure*}[htbp]
\centering
\includegraphics[width=0.35\textwidth]{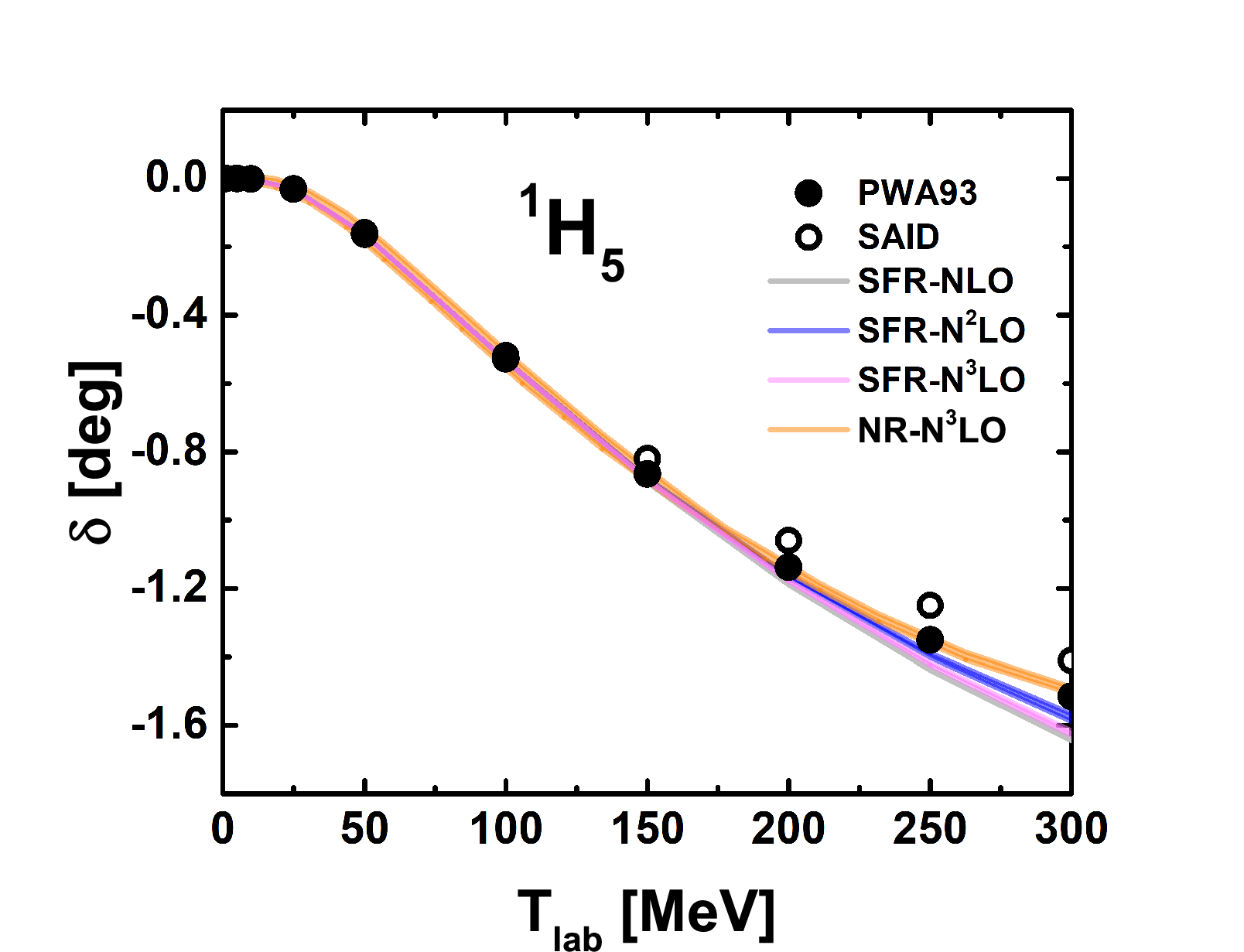}\hspace{-8mm}
\includegraphics[width=0.35\textwidth]{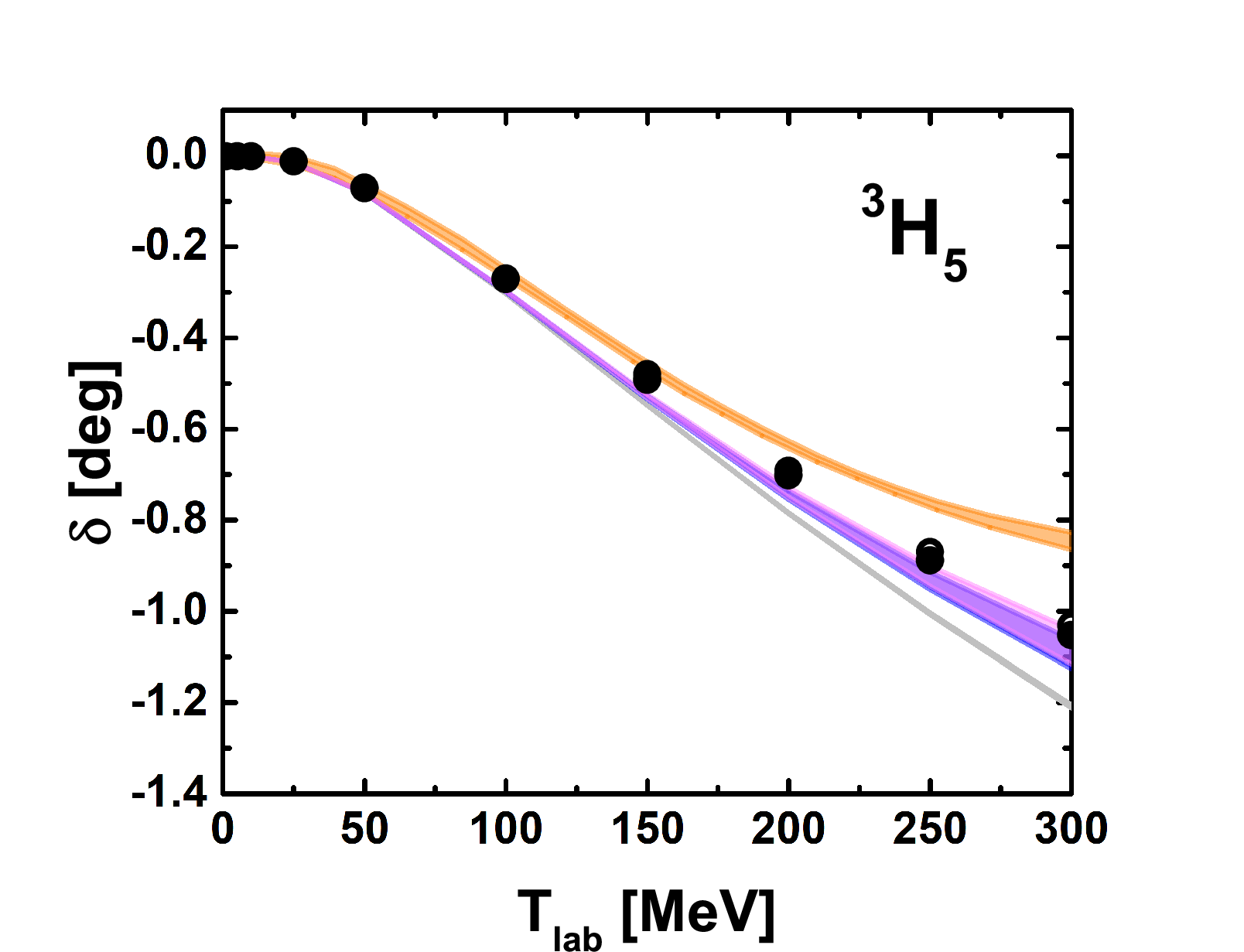}\hspace{-8mm}
\includegraphics[width=0.35\textwidth]{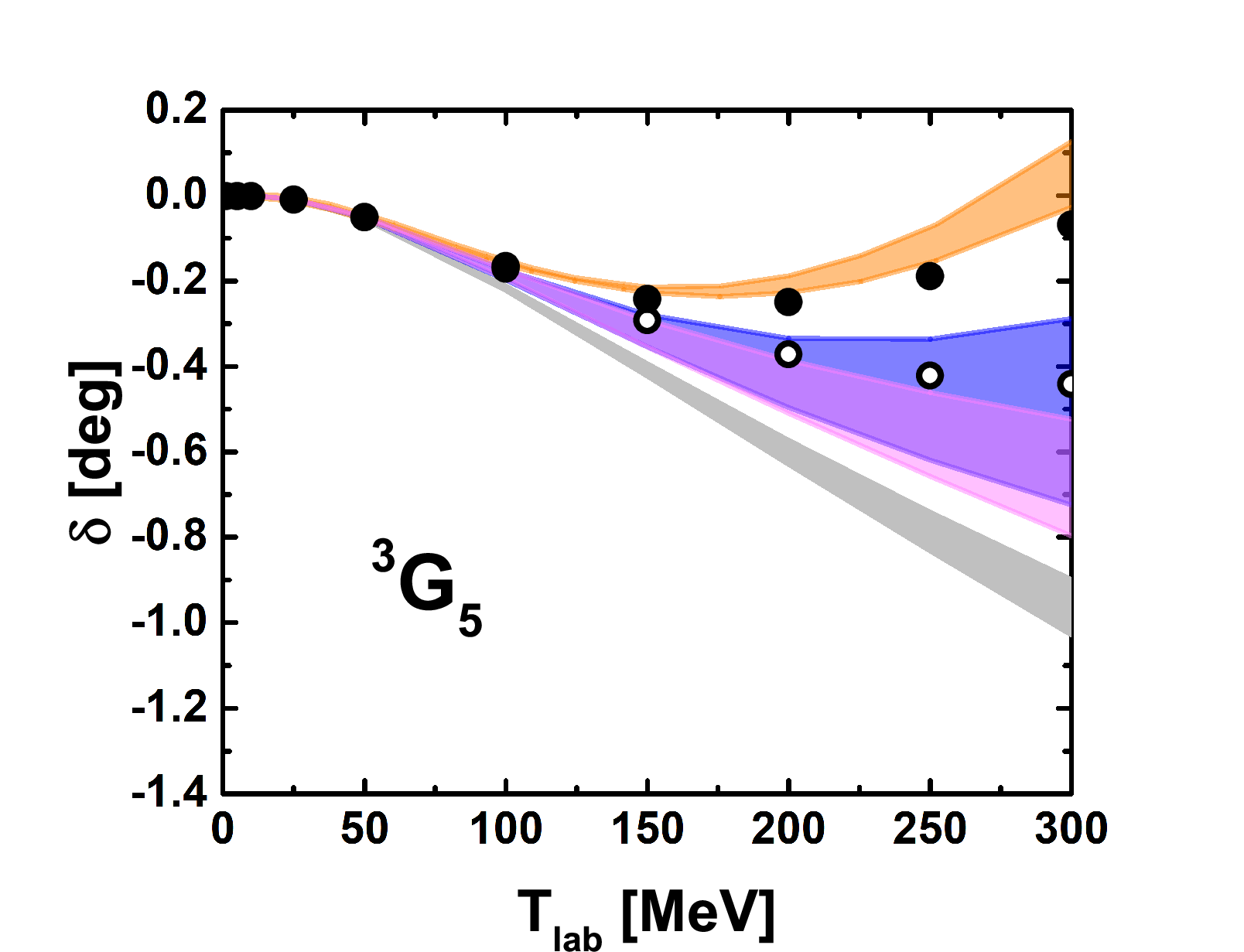}\\ 
\includegraphics[width=0.35\textwidth]{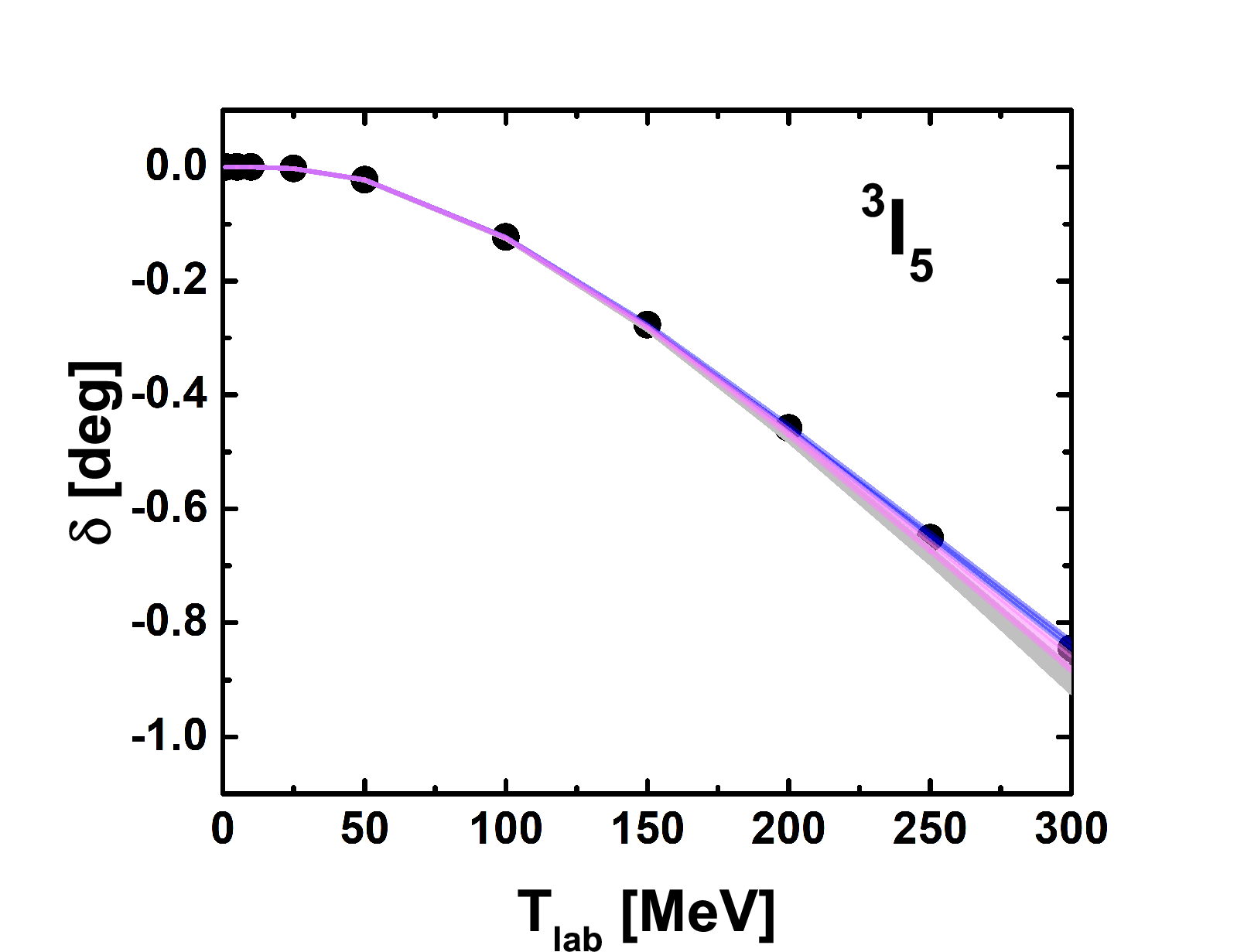}\hspace{-8mm}
\includegraphics[width=0.35\textwidth]{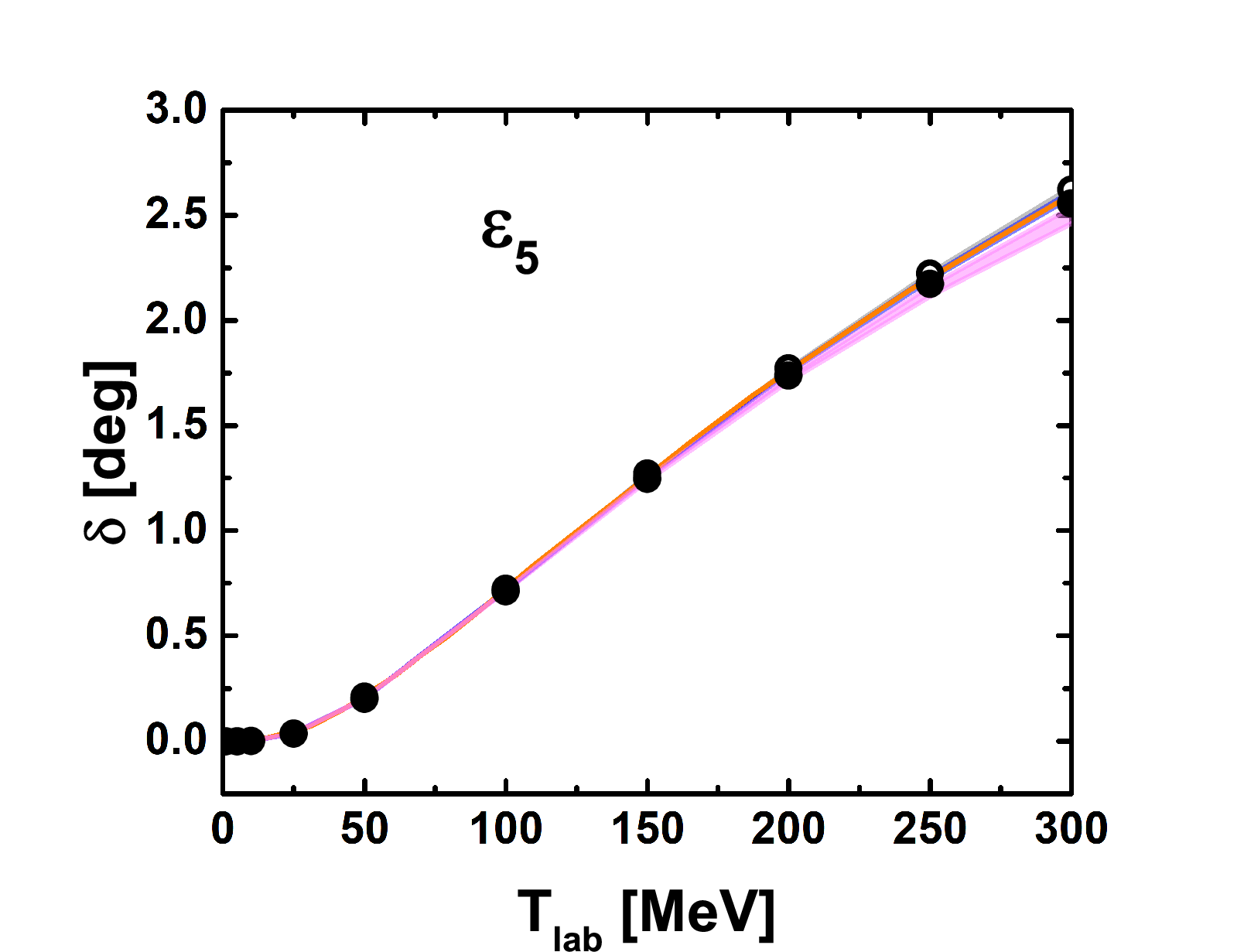}\hspace{-8mm}
\caption{Same as Fig.~\ref{fig:PSJ4} but for the $J=5$ partial waves.}
\label{fig:PSJ5}
\end{figure*}

In Fig.~\ref{fig:PSJ5}, we show the results for the $J=5$ partial waves. For these partial waves, the behavior is dominated by the long-range OPE component, and corrections from higher-order multi-pion exchange components are very small. We can see that the NLO, N$^2$LO, and N$^3$LO results almost overlap in the $^1H_5$, $^3H_5$ partial waves and the mixing angle $\epsilon_5$. A visible improvement only emerges in the higher energy region for the $^3H_5$ and $^3G_5$ partial waves. Note that for these $J=5$ partial waves, the magnitudes of the phase shifts are quite small. Thus, one should be careful when comparing discrepancies between the theoretical results and the two partial-wave analysis results.

\subsection{Comparison with results obtained with dimensional regularization }

\begin{figure*}[htbp]
\centering
\includegraphics[width=0.35\textwidth]{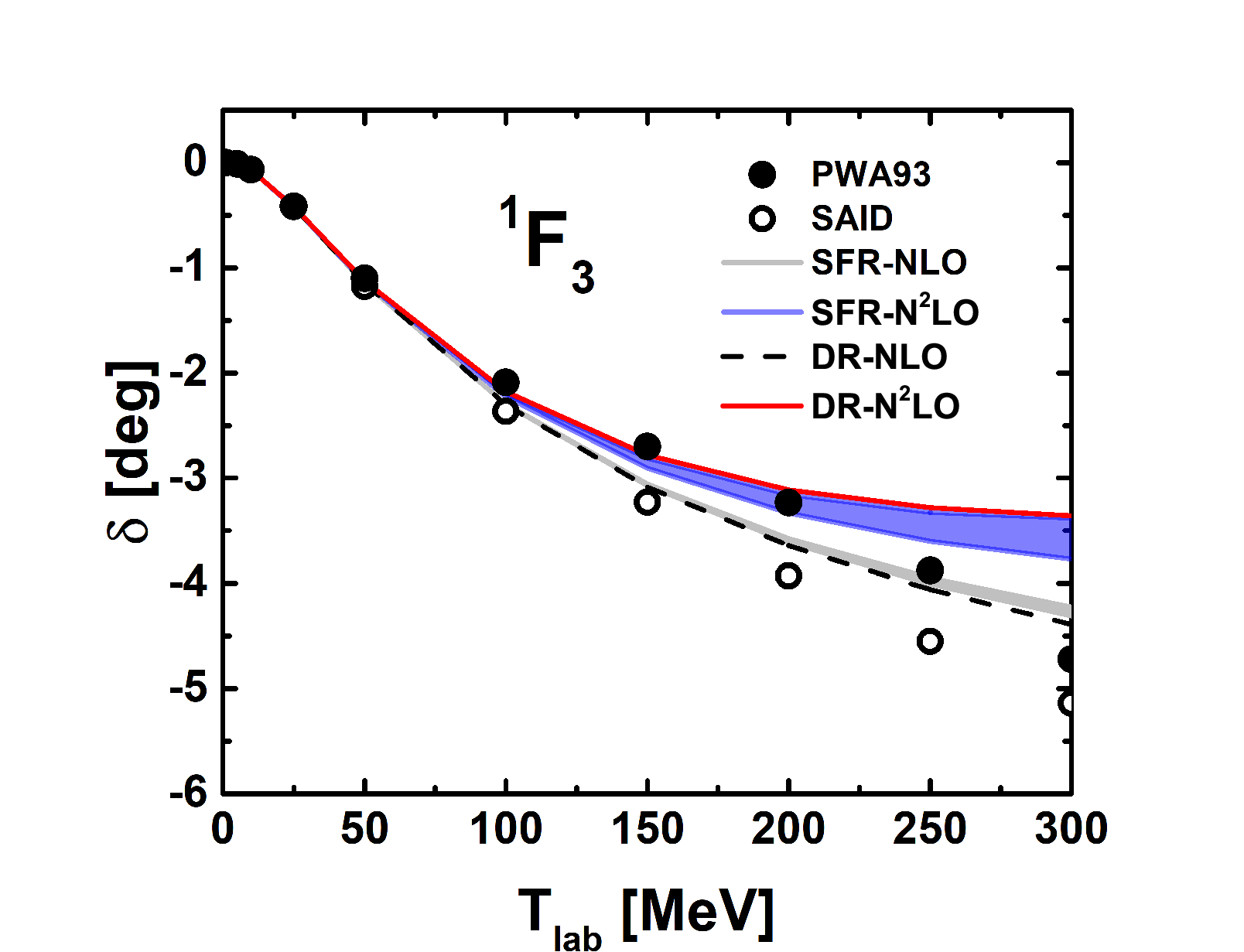}\hspace{-8mm}
\includegraphics[width=0.35\textwidth]{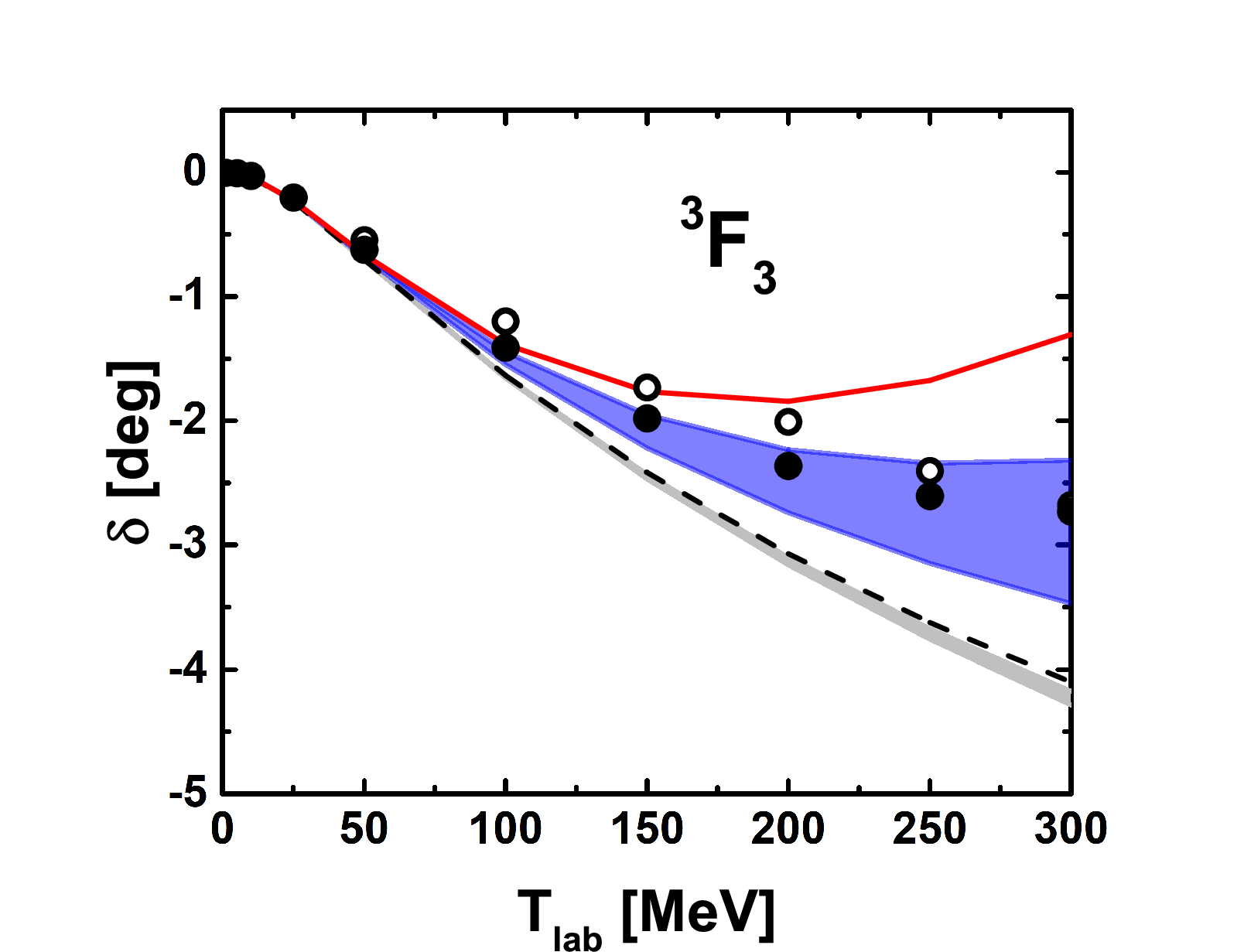}\hspace{-8mm}
\includegraphics[width=0.35\textwidth]{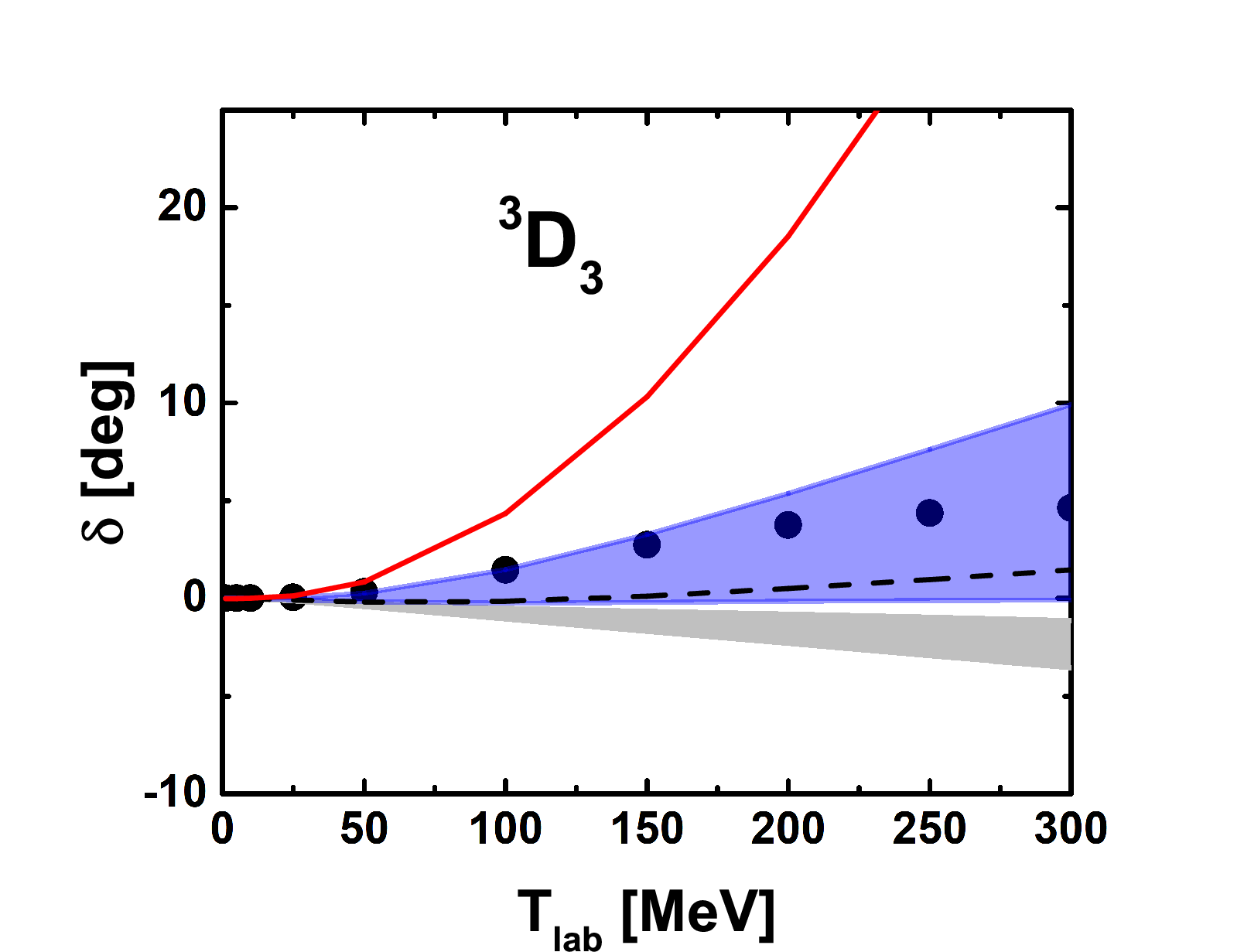}\\ 
\includegraphics[width=0.35\textwidth]{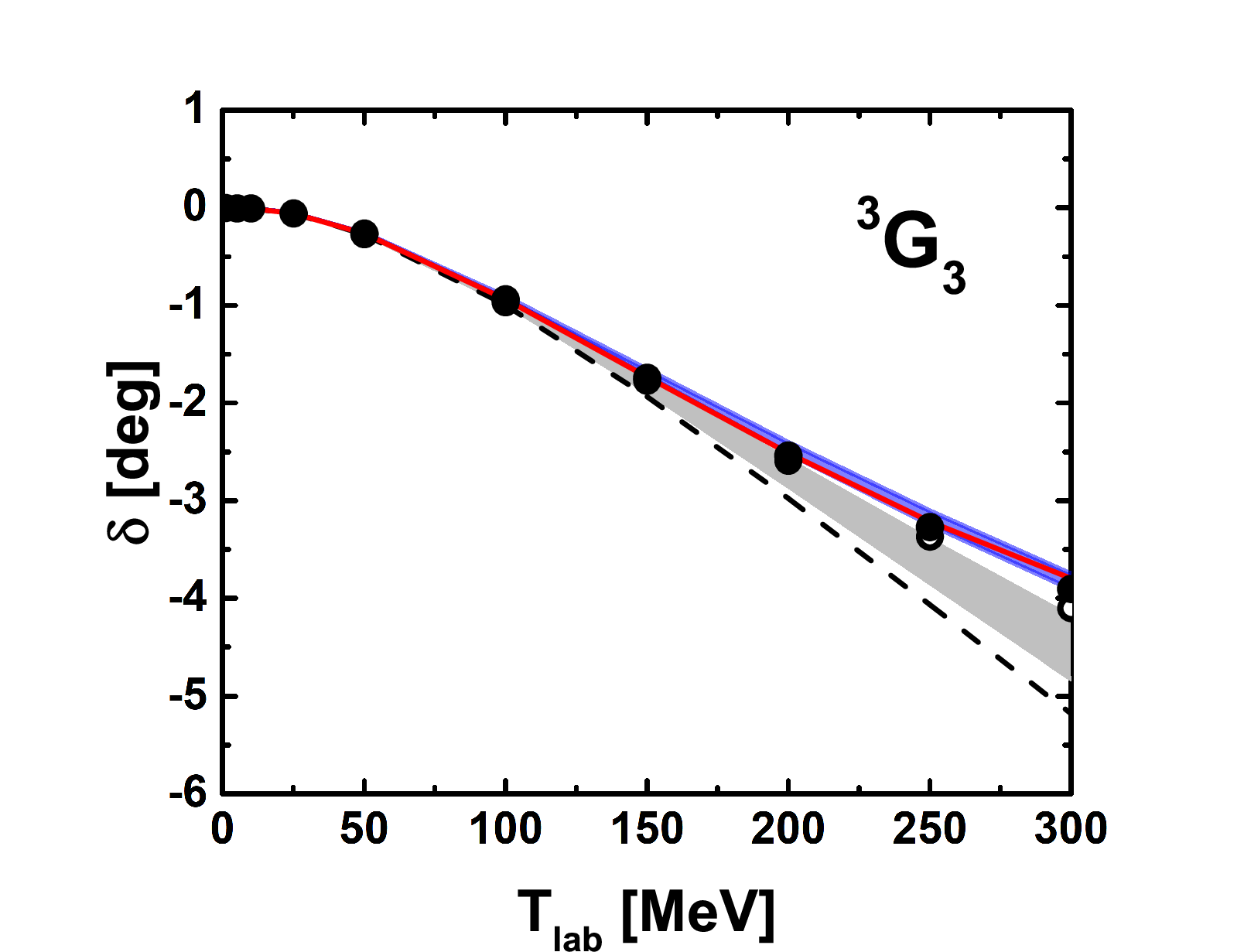}\hspace{-8mm}
\includegraphics[width=0.35\textwidth]{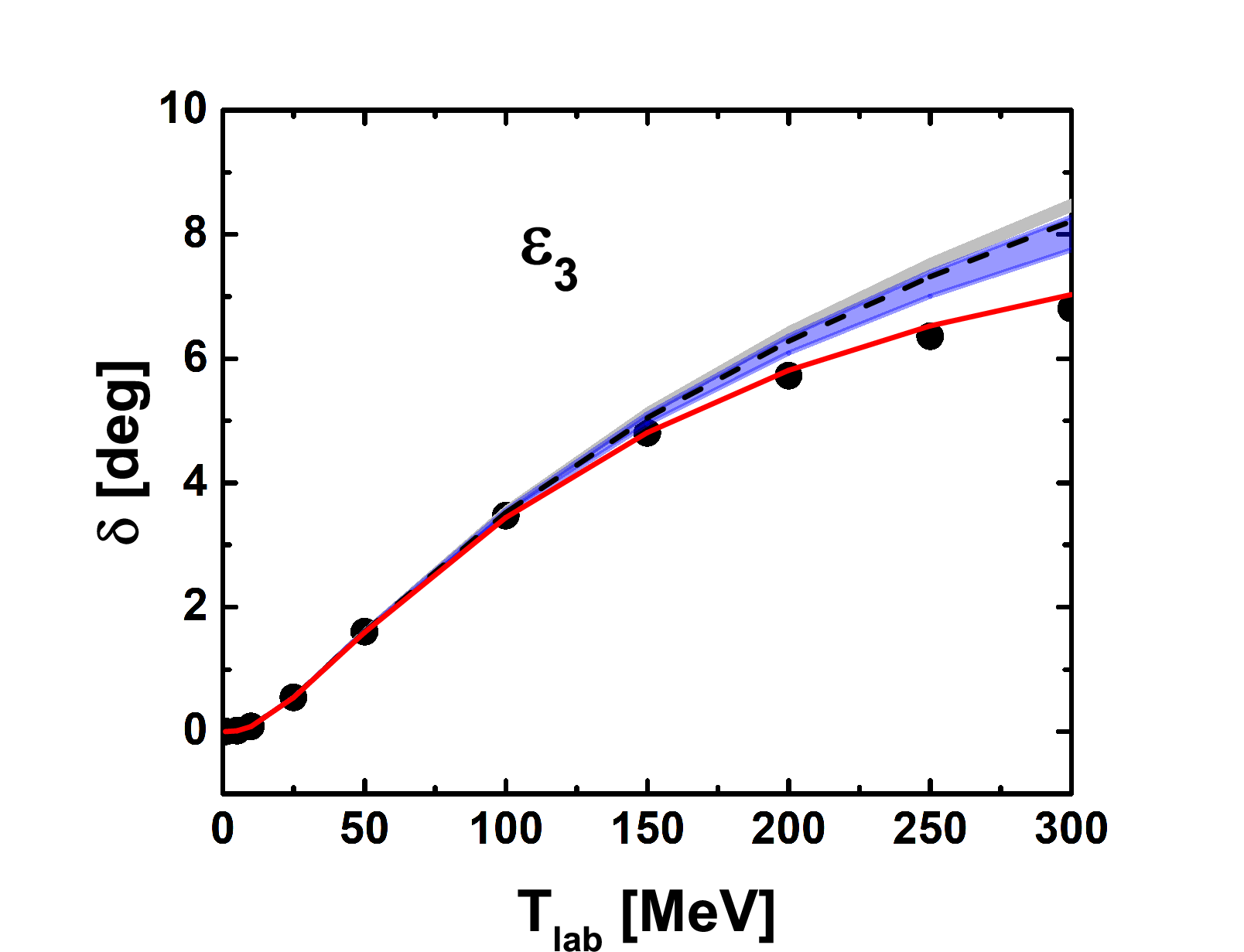}\\ 
\caption{The $J=3$ phase shifts from the spectral function regularization and the dimensional regularization. The gray and blue bands are the results obtained with the SFR method, while the black dashed and solid red lines are those obtained with the DR method~\cite{Xiao:2020ozd}.}
\label{fig:PScpDR3}
\end{figure*}

\begin{figure*}[htbp]
\centering
\includegraphics[width=0.35\textwidth]{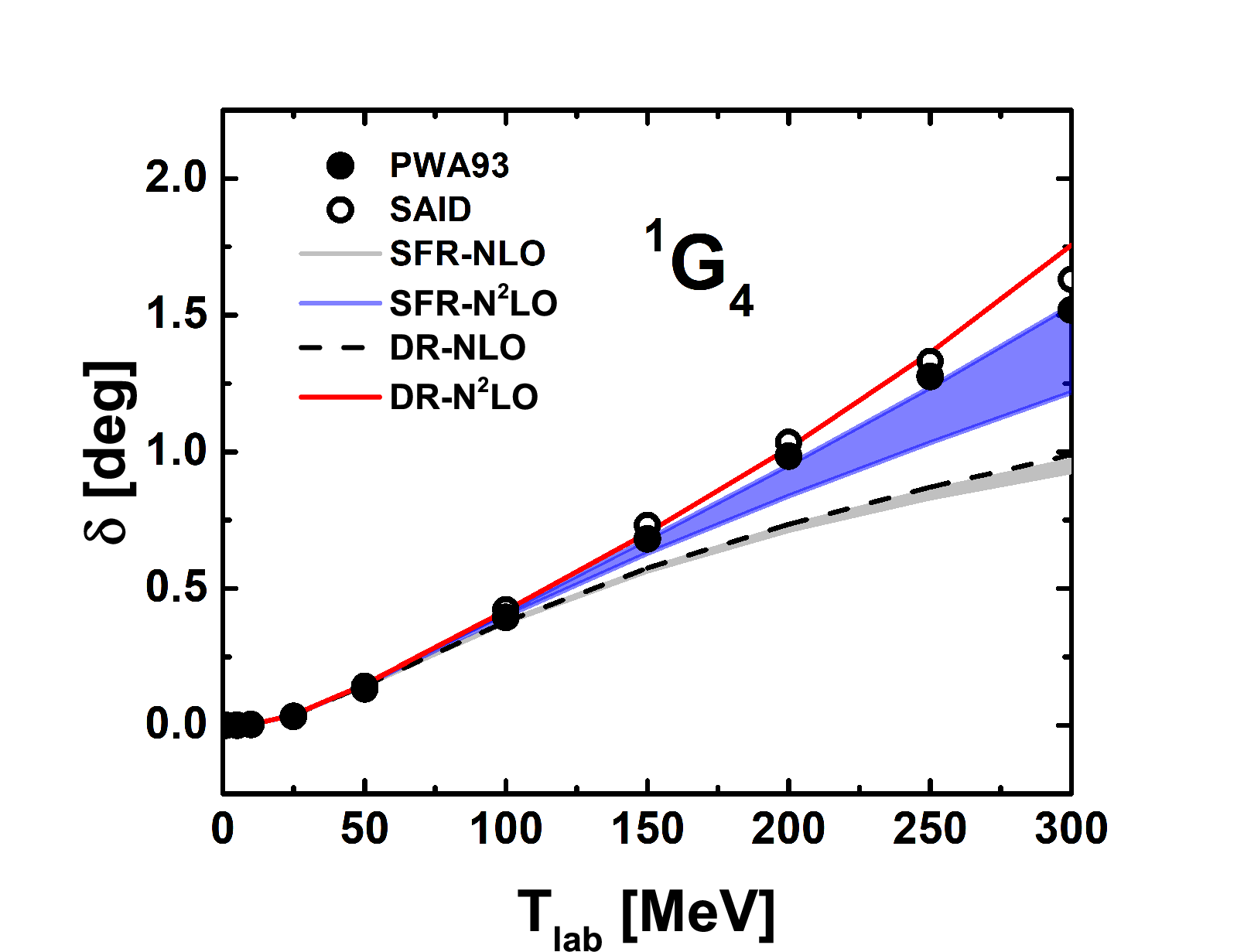}\hspace{-8mm}
\includegraphics[width=0.35\textwidth]{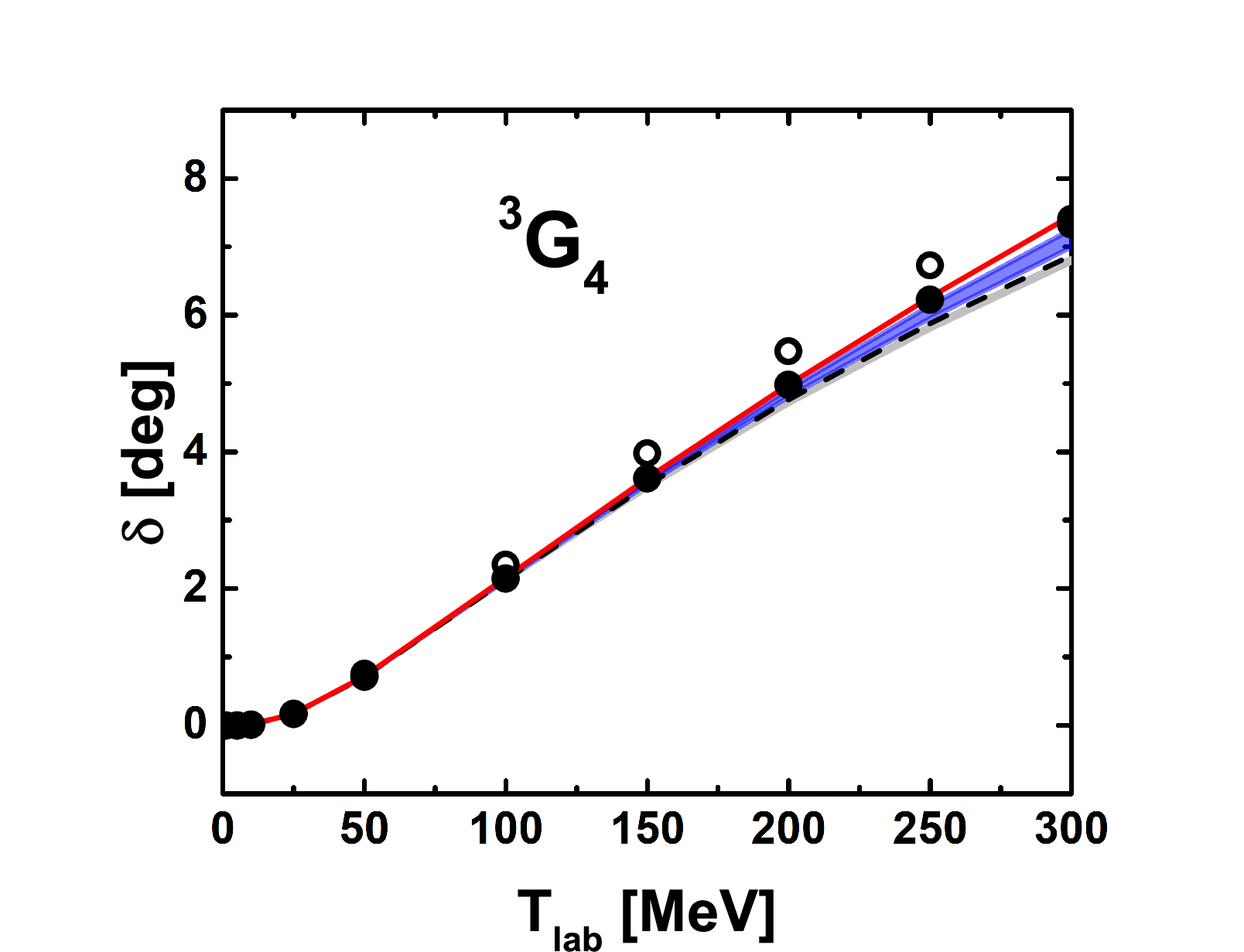}\hspace{-8mm}
\includegraphics[width=0.35\textwidth]{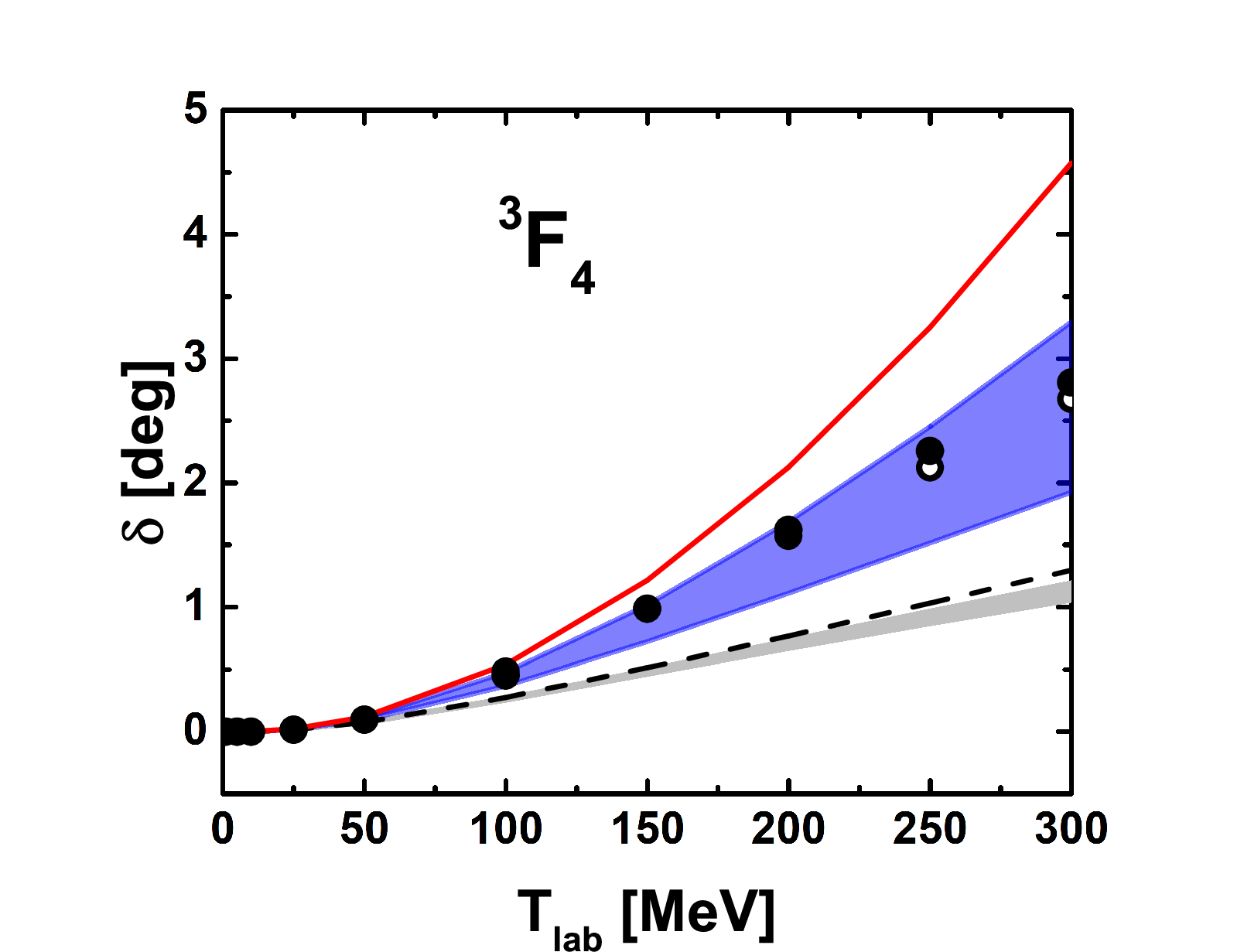}\\ 
\includegraphics[width=0.35\textwidth]{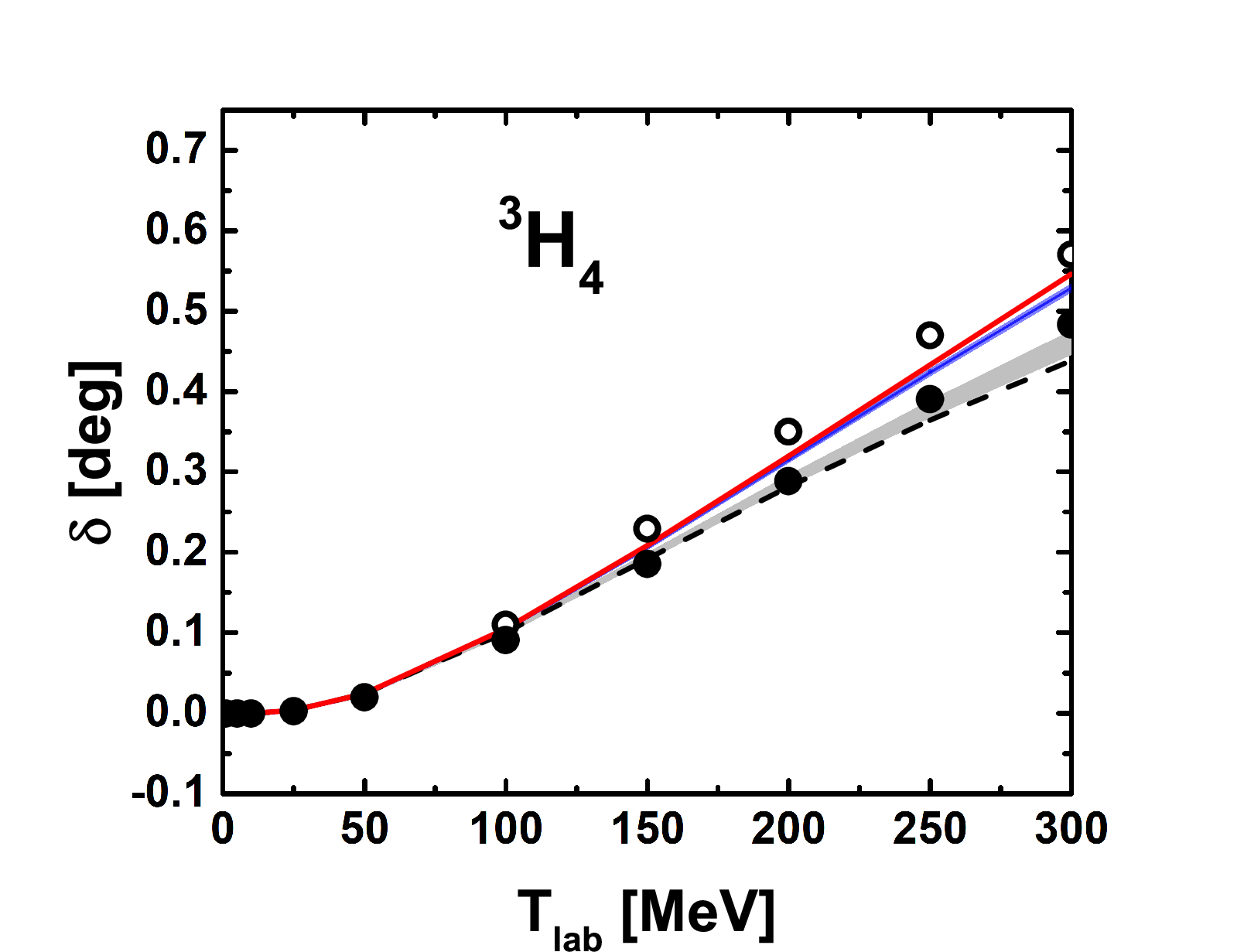}\hspace{-8mm}
\includegraphics[width=0.35\textwidth]{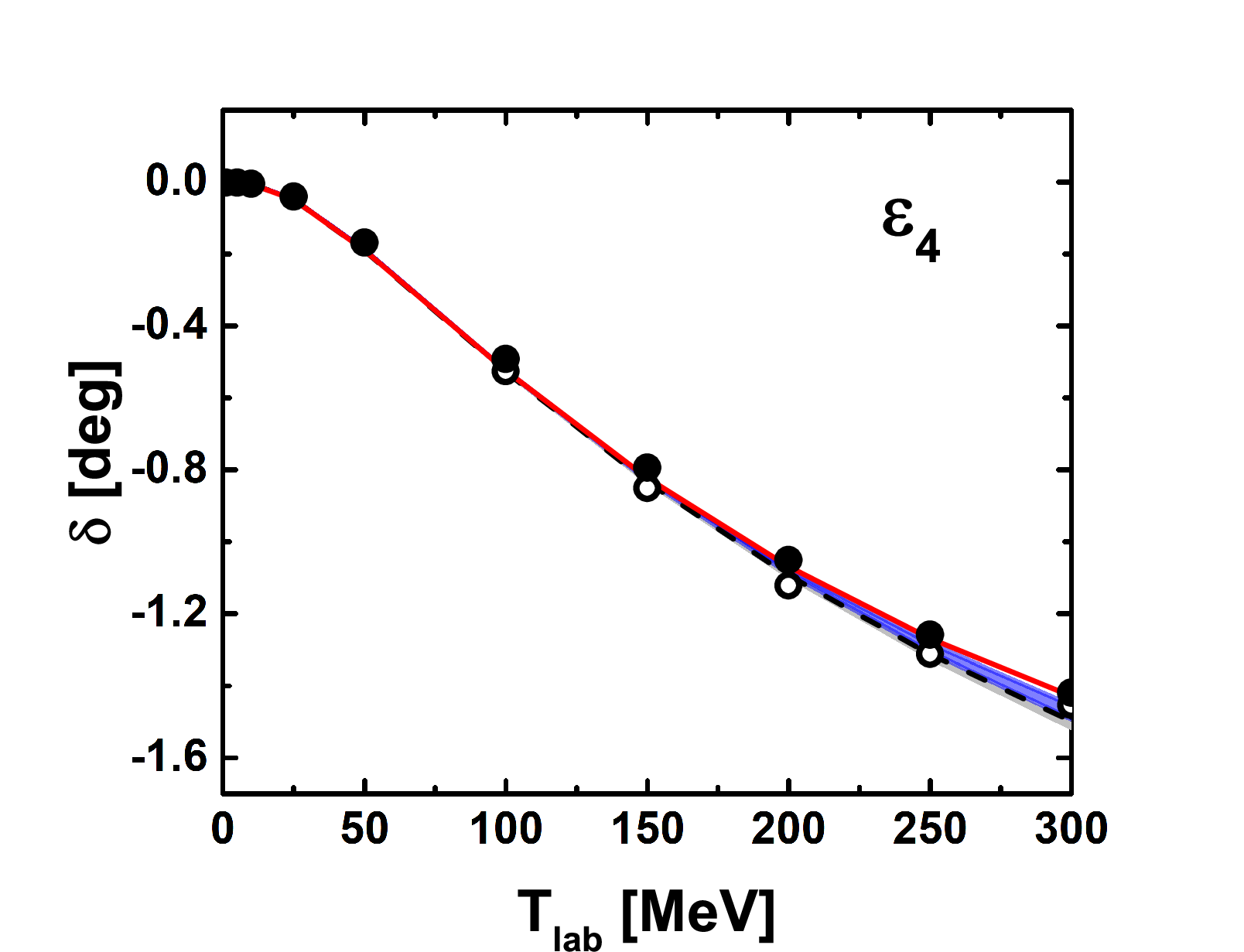}\\ 
\caption{Same as Fig.~\ref{fig:PScpDR3} but for the $J=4$ partial waves.}
\label{fig:PScpDR4}
\end{figure*}

\begin{figure*}[htbp]
\centering
\includegraphics[width=0.35\textwidth]{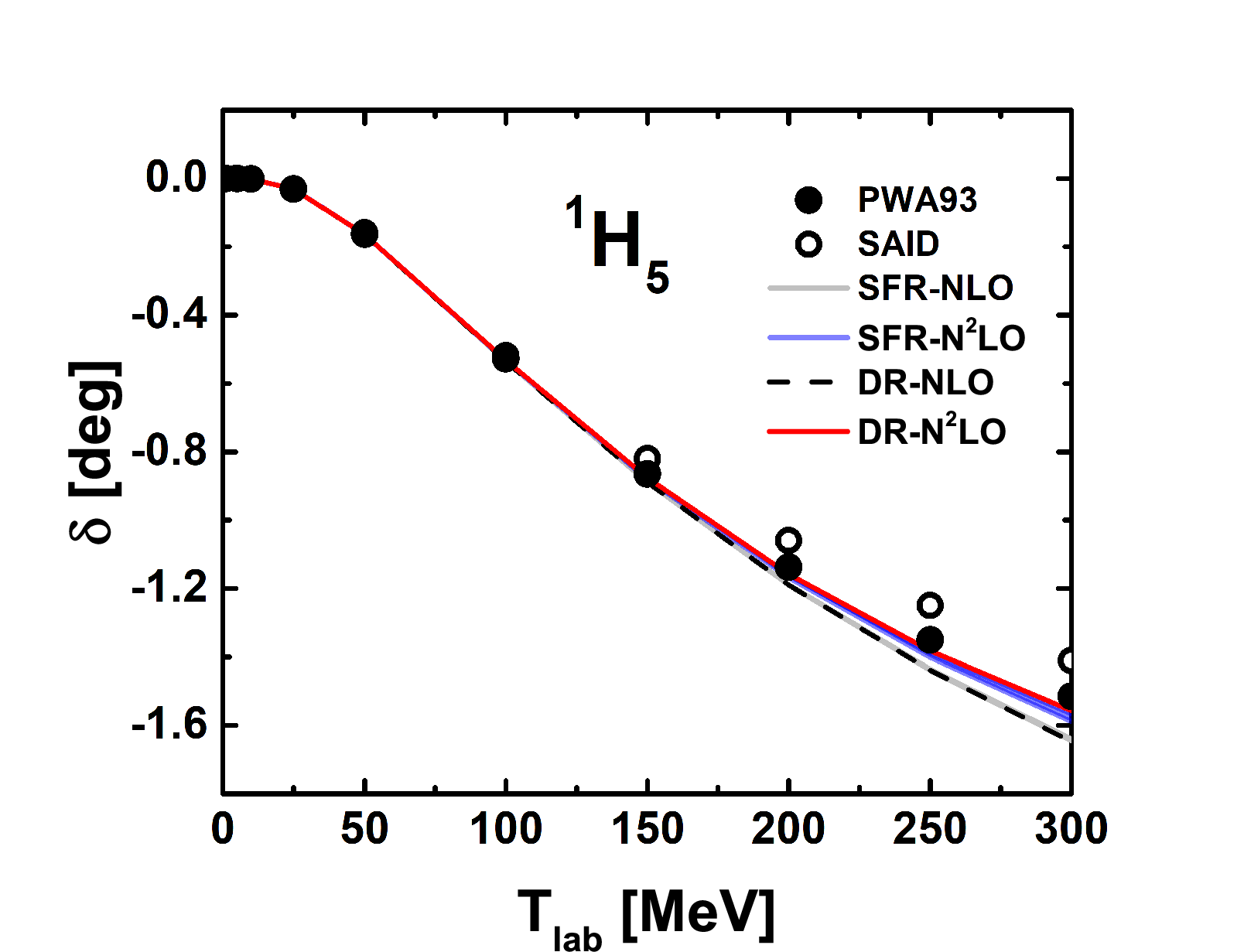}\hspace{-8mm}
\includegraphics[width=0.35\textwidth]{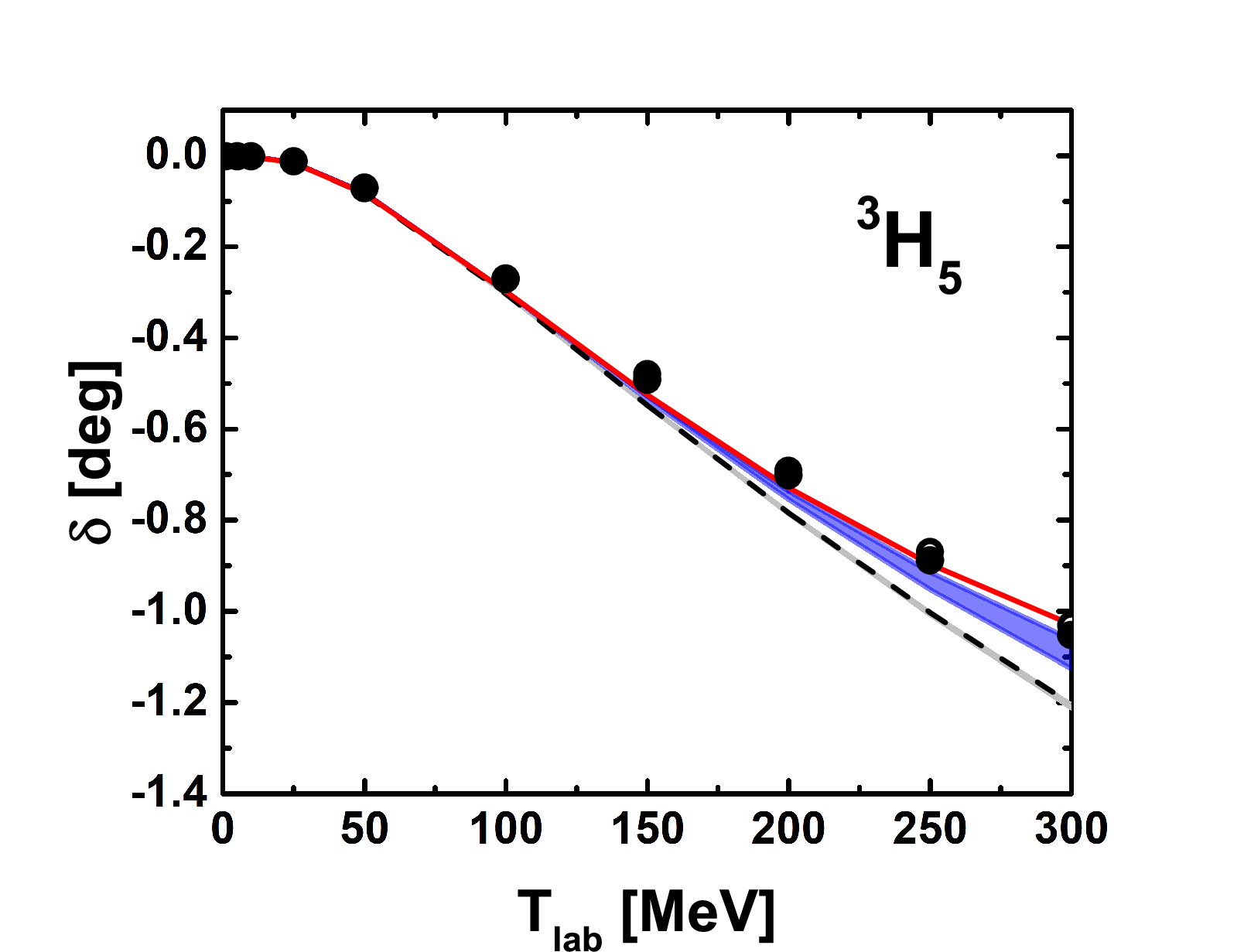}\hspace{-8mm}
\includegraphics[width=0.35\textwidth]{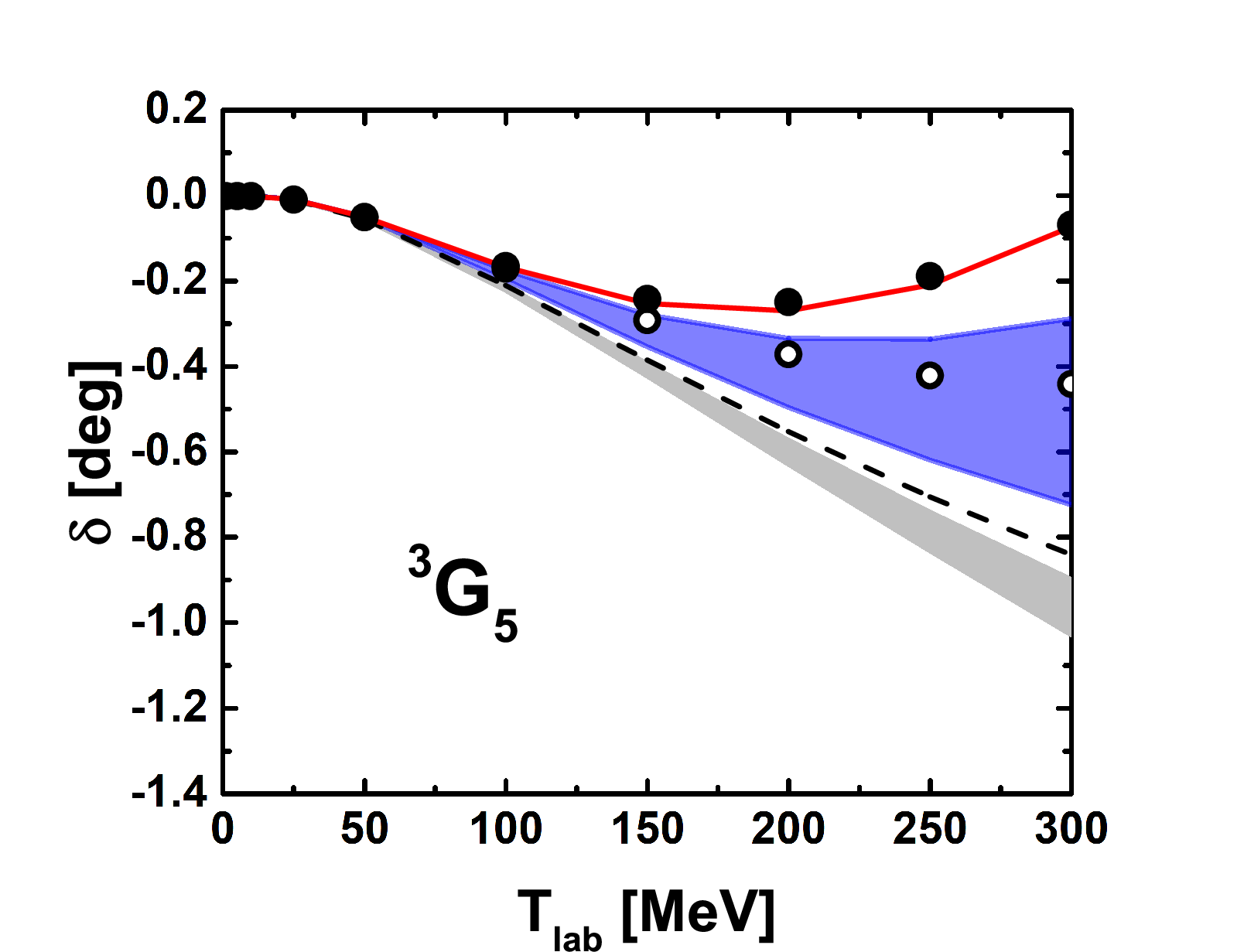}\\ 
\includegraphics[width=0.35\textwidth]{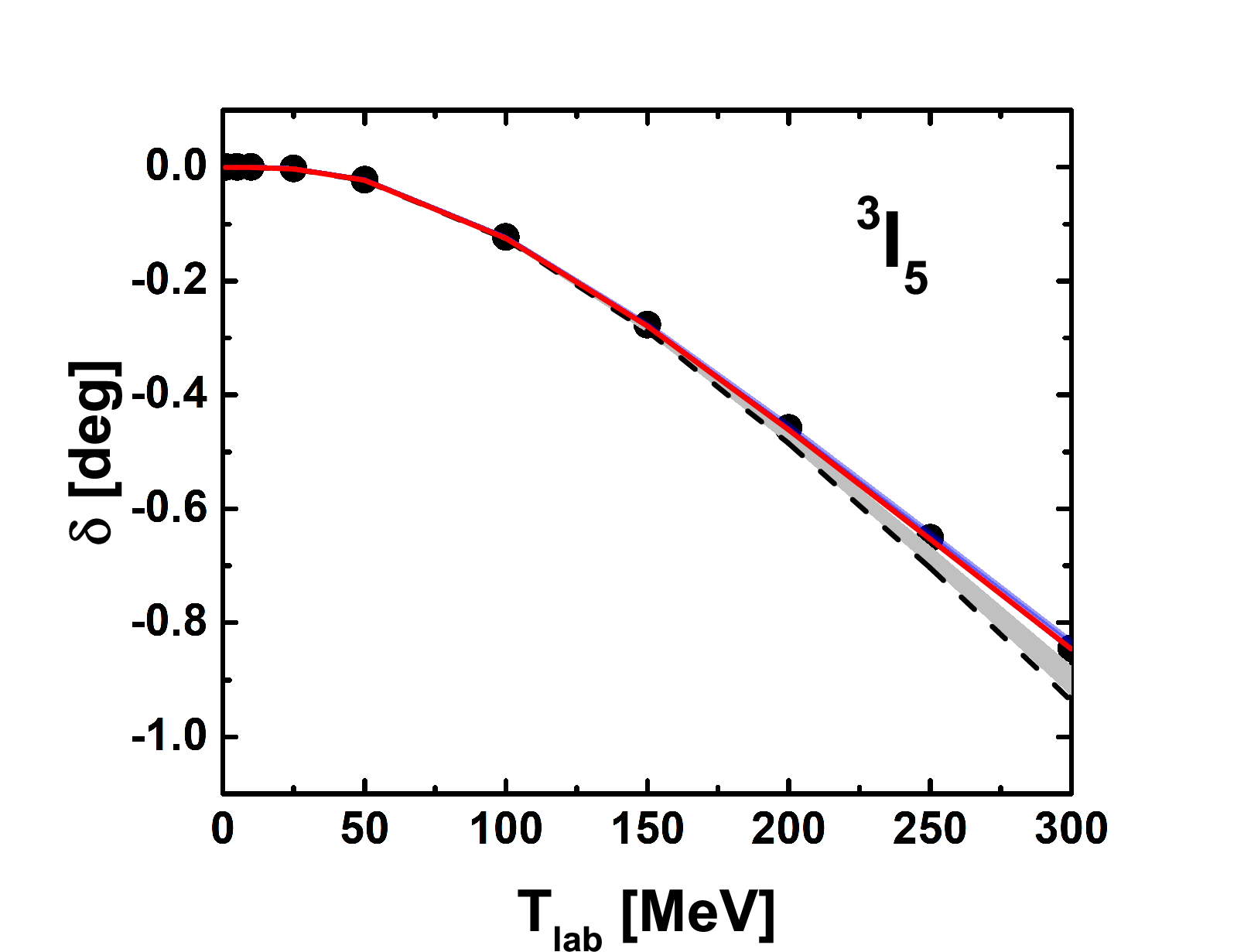}\hspace{-8mm}
\includegraphics[width=0.35\textwidth]{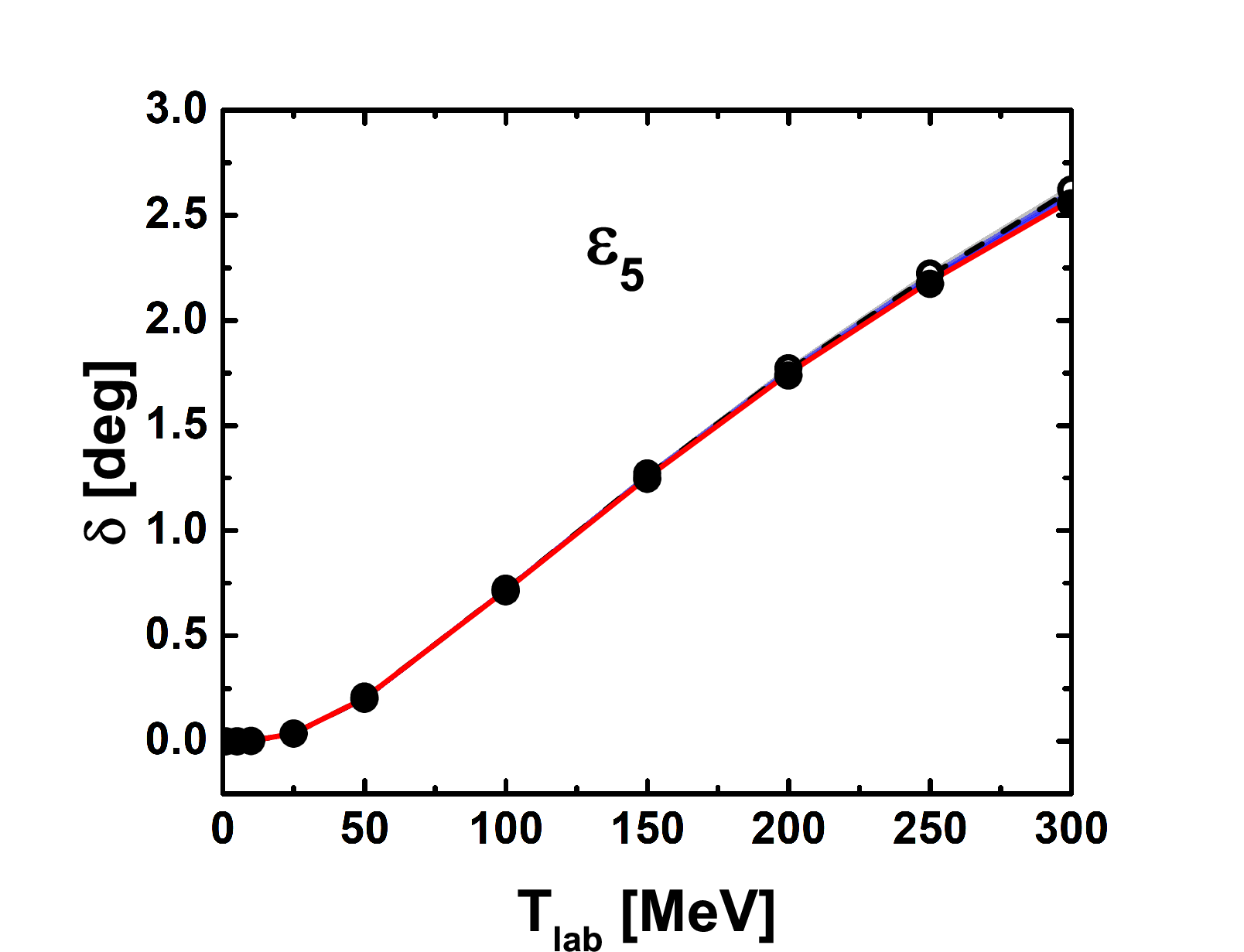}\hspace{-8mm}
\caption{Same as Fig.~\ref{fig:PScpDR3} but for the $J=5$ partial waves.}
\label{fig:PScpDR5}
\end{figure*}

As mentioned above, spectral function regularization is essentially a cutoff regularization.
When the cutoff is taken to infinity, it is identical to dimensional regularization. However, because we introduce a cutoff to constrain the imaginary parts to the low-momentum region where chiral effective field theory is applicable, discrepancies naturally arise between these two methods. In Figs.~\ref{fig:PScpDR3}, \ref{fig:PScpDR4}, and \ref{fig:PScpDR5}, we show such discrepancies between the SFR and DR methods for the NLO and N$^2$LO calculations, which are denoted as ``SFR/DR-NLO" and ``SFR/DR-N$^2$LO". The comparison for the N$^3$LO is absent due to the extreme complexity of two-loop diagrams with the DR method. Note that the results from dimensional regularization are identical to those in our previous work~\cite{Xiao:2020ozd}.

In most peripheral partial waves concerned in the present manuscript, the results from the two methods are highly consistent, except for those with relatively smaller orbital angular momentum $L$, e.g., the $^3F_3$ and $^3F_4$ partial waves. In these two partial waves, it is evident that the corrections from the subleading TPE with the dimensional regularization method are excessively large, leading to overcorrected phase shifts. In contrast, the results from the spectral function regularization method provide more moderate corrections that are in far better agreement with both the PWA93 and SAID data. 

\section{Summary and outlook}\label{sec:Sum}

In the present manuscript, we extended the calculation of two-pion-exchange contributions to the relativistic chiral nuclear force to N$^3$LO. This work is based on the most general covariant baryon chiral effective Lagrangians up to the third chiral order. Given the extreme complexity of the dimensional regularization method for multi-loop calculations, we adopted the spectral-function regularization method for all loop diagrams at NLO, N$^2$LO, and N$^3$LO in our calculations. A Gaussian-type form factor was used to suppress the high-momentum components of the exchanged pions, which are beyond the range of applicability of effective field theories, and the cutoff was varied from $0.5$ to $0.8$ GeV as a rough estimate of the theoretical uncertainties. No significant differences are found when comparing the results of the two regularization methods. Only in the low-angular-momentum partial waves are the results from the spectral function regularization method relatively moderate and in better agreement with the PWA93 or SAID data.

In the peripheral partial waves of interest in the present work, we find that the phase shifts are in very good agreement with the PWA93 and SAID data. What is particularly worth highlighting is that in most partial waves, such a good agreement is already achieved up to N$^2$LO and the visible corrections from the fourth chiral order (N$^3$LO) only emerge in very few partial waves with smaller orbital angular momentum $L$ in the higher energy region, e.g., $^1F_3$. Compared to the non-relativistic calculations where the fifth chiral order (N$^4$LO) contributions are still necessary to provide repulsive interactions to compensate for the excessive attraction from the lower chiral orders~\cite{Entem:2014msa}, such a pattern indicates a faster convergence of the relativistic chiral nuclear force. We can conclude that, at least for these peripheral partial waves, even the N$^3$LO loop contributions may not be necessary in the relativistic framework.

In the next stage, we will pin down the LECs needed for the N$^3$LO $NN$ forces by the phase shifts of relevant partial waves and examine the description of $NN$ scattering observables of the relativistic N$^3$LO chiral nuclear force. Such studies are of great relevance for nuclear structure and reaction studies in the relativistic framework~\cite{Zou:2023quo,Zheng:2025sol,Zou:2025dao,Shen:2025iue}.

\section{Acknowledgment}
JXL thanks Prof. Norbert Kaiser for enlightening communications regarding non-relativistic two-loop contributions.
This work is supported in part by the National Natural Science Foundation of China under Grants Nos. 12435007 and 1252200936.

\bibliography{refs}

@article{Chen:2024twu,
    author = "Chen, Long-Bin and Hu, Siwei and Jia, Yu and Mo, Zhewen",
    title = "{Light quark mass dependence of nucleon mass to two-loop order}",
    eprint = "2406.04124",
    archivePrefix = "arXiv",
    primaryClass = "hep-ph",
    month = "6",
    year = "2024"
}

@article{Ledwig:2014rfa,
    author = "Ledwig, T. and Martin Camalich, J. and Geng, L. S. and Vicente Vacas, M. J.",
    title = "{Octet-baryon axial-vector charges and SU(3)-breaking effects in the semileptonic hyperon decays}",
    eprint = "1405.5456",
    archivePrefix = "arXiv",
    primaryClass = "hep-ph",
    doi = "10.1103/PhysRevD.90.054502",
    journal = "Phys. Rev. D",
    volume = "90",
    number = "5",
    pages = "054502",
    year = "2014"
}

@article{Hagelstein:2015egb,
    author = "Hagelstein, Franziska and Miskimen, Rory and Pascalutsa, Vladimir",
    title = "{Nucleon Polarizabilities: from Compton Scattering to Hydrogen Atom}",
    eprint = "1512.03765",
    archivePrefix = "arXiv",
    primaryClass = "nucl-th",
    reportNumber = "MITP-15-119",
    doi = "10.1016/j.ppnp.2015.12.001",
    journal = "Prog. Part. Nucl. Phys.",
    volume = "88",
    pages = "29--97",
    year = "2016"
}

@article{Lensky:2016nui,
    author = "Lensky, Vadim and Pascalutsa, Vladimir and Vanderhaeghen, Marc",
    title = "{Generalized polarizabilities of the nucleon in baryon chiral perturbation theory}",
    eprint = "1612.08626",
    archivePrefix = "arXiv",
    primaryClass = "hep-ph",
    reportNumber = "MITP-16-118",
    doi = "10.1140/epjc/s10052-017-4652-9",
    journal = "Eur. Phys. J. C",
    volume = "77",
    number = "2",
    pages = "119",
    year = "2017"
}

@article{Yao:2018pzc,
    author = "Yao, De-Liang and Alvarez-Ruso, Luis and Hiller Blin, Astrid N. and Vicente Vacas, M. J.",
    title = "{Weak pion production off the nucleon in covariant chiral perturbation theory}",
    eprint = "1806.09364",
    archivePrefix = "arXiv",
    primaryClass = "hep-ph",
    doi = "10.1103/PhysRevD.98.076004",
    journal = "Phys. Rev. D",
    volume = "98",
    number = "7",
    pages = "076004",
    year = "2018"
}

@article{Yao:2019avf,
    author = "Yao, De-Liang and Alvarez-Ruso, Luis and Vicente Vacas, M. J.",
    title = "{Neutral-current weak pion production off the nucleon in covariant chiral perturbation theory}",
    eprint = "1901.00773",
    archivePrefix = "arXiv",
    primaryClass = "hep-ph",
    doi = "10.1016/j.physletb.2019.05.036",
    journal = "Phys. Lett. B",
    volume = "794",
    pages = "109--113",
    year = "2019"
}

@article{Zheng:2025sol,
    author = "Zheng, Ru-You and Liu, Zhi-Wei and Geng, Li-Sheng and Hu, Jin-Niu and Wang, Sibo",
    title = "{In-medium {\ensuremath{\Lambda}}N interactions with leading order covariant chiral hyperon/nucleon-nucleon forces}",
    eprint = "2501.02826",
    archivePrefix = "arXiv",
    primaryClass = "nucl-th",
    doi = "10.1016/j.physletb.2025.139416",
    journal = "Phys. Lett. B",
    volume = "864",
    pages = "139416",
    year = "2025"
}

@article{Shen:2025iue,
    author = "Shen, Shihang and Lu, Jun-Xu and Geng, Li-Sheng and Meng, Jie and Zou, Wei-Jiang",
    title = "{From bare two-nucleon interaction to nuclear matter and finite nuclei in a relativistic framework}",
    eprint = "2507.01257",
    archivePrefix = "arXiv",
    primaryClass = "nucl-th",
    month = "7",
    year = "2025"
}

@article{Zou:2025dao,
    author = "Zou, Wei-Jiang and Yang, Yi-Long and Lu, Jun-Xu and Zhao, Peng-Wei and Geng, Li-Sheng and Meng, Jie",
    title = "{Nuclear and neutron matter in the relativistic Brueckner-Hartree-Fock theory with next-to-leading order covariant chiral nuclear force}",
    eprint = "2506.18519",
    archivePrefix = "arXiv",
    primaryClass = "nucl-th",
    month = "6",
    year = "2025"
}

@article{Zou:2023quo,
    author = "Zou, Wei-Jiang and Lu, Jun-Xu and Zhao, Peng-Wei and Geng, Li-Sheng and Meng, Jie",
    title = "{Saturation of nuclear matter in the relativistic Brueckner-Hatree-Fock approach with a leading order covariant chiral nuclear force}",
    eprint = "2312.15672",
    archivePrefix = "arXiv",
    primaryClass = "nucl-th",
    doi = "10.1016/j.physletb.2024.138732",
    journal = "Phys. Lett. B",
    volume = "854",
    pages = "138732",
    year = "2024"
}

@article{Geng:2013xn,
    author = "Geng, Lisheng",
    title = "{Recent developments in SU(3) covariant baryon chiral perturbation theory}",
    eprint = "1301.6815",
    archivePrefix = "arXiv",
    primaryClass = "nucl-th",
    doi = "10.1007/s11467-013-0327-7",
    journal = "Front. Phys. (Beijing)",
    volume = "8",
    pages = "328--348",
    year = "2013"
}

@article{Shi:2022dhw,
    author = "Shi, Rui-Xiang and Li, Shuang-Yi and Lu, Jun-Xu and Geng, Li-Sheng",
    title = "{Weak radiative hyperon decays in covariant baryon chiral perturbation theory}",
    eprint = "2206.11773",
    archivePrefix = "arXiv",
    primaryClass = "hep-ph",
    doi = "10.1016/j.scib.2022.10.026",
    journal = "Sci. Bull.",
    volume = "67",
    pages = "2298--2304",
    year = "2022"
}

@article{Ren:2017fbv,
    author = "Ren, Xiu-Lei and Ling, Xi-Zhe and Geng, Li-Sheng",
    title = "{Pion\textendash{}nucleon sigma term revisited in covariant baryon chiral perturbation theory}",
    eprint = "1710.07164",
    archivePrefix = "arXiv",
    primaryClass = "hep-ph",
    doi = "10.1016/j.physletb.2018.05.063",
    journal = "Phys. Lett. B",
    volume = "783",
    pages = "7--12",
    year = "2018"
}

@article{Wiringa:1994wb,
    author = "Wiringa, Robert B. and Stoks, V. G. J. and Schiavilla, R.",
    title = "{An Accurate nucleon-nucleon potential with charge independence breaking}",
    eprint = "nucl-th/9408016",
    archivePrefix = "arXiv",
    reportNumber = "PHY-7742-TH-94, CEBAF-TH-94-19",
    doi = "10.1103/PhysRevC.51.38",
    journal = "Phys. Rev. C",
    volume = "51",
    pages = "38--51",
    year = "1995"
}

@article{Machleidt:2000ge,
    author = "Machleidt, R.",
    title = "{The High precision, charge dependent Bonn nucleon-nucleon potential (CD-Bonn)}",
    eprint = "nucl-th/0006014",
    archivePrefix = "arXiv",
    doi = "10.1103/PhysRevC.63.024001",
    journal = "Phys. Rev. C",
    volume = "63",
    pages = "024001",
    year = "2001"
}

@article{Epelbaum:2008ga,
    author = "Epelbaum, Evgeny and Hammer, Hans-Werner and Meissner, Ulf-G.",
    title = "{Modern Theory of Nuclear Forces}",
    eprint = "0811.1338",
    archivePrefix = "arXiv",
    primaryClass = "nucl-th",
    reportNumber = "HISKP-TH-08-18, FZJ-IKP-TH-2008-20",
    doi = "10.1103/RevModPhys.81.1773",
    journal = "Rev. Mod. Phys.",
    volume = "81",
    pages = "1773--1825",
    year = "2009"
}

@article{Weinberg:1990rz,
    author = "Weinberg, Steven",
    title = "{Nuclear forces from chiral Lagrangians}",
    reportNumber = "UTTG-31-90",
    doi = "10.1016/0370-2693(90)90938-3",
    journal = "Phys. Lett. B",
    volume = "251",
    pages = "288--292",
    year = "1990"
}

@article{Kaiser:1997mw,
    author = "Kaiser, Norbert and Brockmann, R. and Weise, W.",
    title = "{Peripheral nucleon-nucleon phase shifts and chiral symmetry}",
    eprint = "nucl-th/9706045",
    archivePrefix = "arXiv",
    doi = "10.1016/S0375-9474(97)00586-1",
    journal = "Nucl. Phys. A",
    volume = "625",
    pages = "758--788",
    year = "1997"
}

@article{Epelbaum:1999dj,
    author = "Epelbaum, E. and Gloeckle, Walter and Meissner, Ulf-G.",
    title = "{Nuclear forces from chiral Lagrangians using the method of unitary transformation. 2. The two nucleon system}",
    eprint = "nucl-th/9910064",
    archivePrefix = "arXiv",
    reportNumber = "FZJ-IKP-TH-1999-19, FZJ-IKP(TH)-1999-19",
    doi = "10.1016/S0375-9474(99)00821-0",
    journal = "Nucl. Phys. A",
    volume = "671",
    pages = "295--331",
    year = "2000"
}

@article{Epelbaum:2003gr,
    author = "Epelbaum, Evgeny and Gloeckle, Walter and Meissner, Ulf-G.",
    title = "{Improving the convergence of the chiral expansion for nuclear forces. 1. Peripheral phases}",
    eprint = "nucl-th/0304037",
    archivePrefix = "arXiv",
    doi = "10.1140/epja/i2003-10096-0",
    journal = "Eur. Phys. J. A",
    volume = "19",
    pages = "125--137",
    year = "2004"
}

@article{Entem:2003ft,
    author = "Entem, D. R. and Machleidt, R.",
    title = "{Accurate charge dependent nucleon nucleon potential at fourth order of chiral perturbation theory}",
    eprint = "nucl-th/0304018",
    archivePrefix = "arXiv",
    doi = "10.1103/PhysRevC.68.041001",
    journal = "Phys. Rev. C",
    volume = "68",
    pages = "041001",
    year = "2003"
}

@article{Epelbaum:2014efa,
    author = "Epelbaum, E. and Krebs, H. and Mei\ss{}ner, U. G.",
    title = "{Improved chiral nucleon-nucleon potential up to next-to-next-to-next-to-leading order}",
    eprint = "1412.0142",
    archivePrefix = "arXiv",
    primaryClass = "nucl-th",
    doi = "10.1140/epja/i2015-15053-8",
    journal = "Eur. Phys. J. A",
    volume = "51",
    number = "5",
    pages = "53",
    year = "2015"
}

@article{Saha:2022oep,
    author = "Saha, S. K. and Entem, D. R. and Machleidt, R. and Nosyk, Y.",
    title = "{Local position-space two-nucleon potentials from leading to fourth order of chiral effective field theory}",
    eprint = "2209.13170",
    archivePrefix = "arXiv",
    primaryClass = "nucl-th",
    doi = "10.1103/PhysRevC.107.034002",
    journal = "Phys. Rev. C",
    volume = "107",
    number = "3",
    pages = "034002",
    year = "2023"
}

@article{Entem:2014msa,
    author = "Entem, D. R. and Kaiser, N. and Machleidt, R. and Nosyk, Y.",
    title = "{Peripheral nucleon-nucleon scattering at fifth order of chiral perturbation theory}",
    eprint = "1411.5335",
    archivePrefix = "arXiv",
    primaryClass = "nucl-th",
    doi = "10.1103/PhysRevC.91.014002",
    journal = "Phys. Rev. C",
    volume = "91",
    number = "1",
    pages = "014002",
    year = "2015"
}

@article{Entem:2017gor,
    author = "Entem, D. R. and Machleidt, R. and Nosyk, Y.",
    title = "{High-quality two-nucleon potentials up to fifth order of the chiral expansion}",
    eprint = "1703.05454",
    archivePrefix = "arXiv",
    primaryClass = "nucl-th",
    doi = "10.1103/PhysRevC.96.024004",
    journal = "Phys. Rev. C",
    volume = "96",
    number = "2",
    pages = "024004",
    year = "2017"
}

@article{Epelbaum:2014sza,
    author = "Epelbaum, E. and Krebs, H. and Mei\ss{}ner, U. G.",
    title = "{Precision nucleon-nucleon potential at fifth order in the chiral expansion}",
    eprint = "1412.4623",
    archivePrefix = "arXiv",
    primaryClass = "nucl-th",
    doi = "10.1103/PhysRevLett.115.122301",
    journal = "Phys. Rev. Lett.",
    volume = "115",
    number = "12",
    pages = "122301",
    year = "2015"
}

@article{Reinert:2017usi,
    author = "Reinert, P. and Krebs, H. and Epelbaum, E.",
    title = "{Semilocal momentum-space regularized chiral two-nucleon potentials up to fifth order}",
    eprint = "1711.08821",
    archivePrefix = "arXiv",
    primaryClass = "nucl-th",
    doi = "10.1140/epja/i2018-12516-4",
    journal = "Eur. Phys. J. A",
    volume = "54",
    number = "5",
    pages = "86",
    year = "2018"
}

@article{Entem:2015xwa,
    author = "Entem, D. R. and Kaiser, N. and Machleidt, R. and Nosyk, Y.",
    title = "{Dominant contributions to the nucleon-nucleon interaction at sixth order of chiral perturbation theory}",
    eprint = "1505.03562",
    archivePrefix = "arXiv",
    primaryClass = "nucl-th",
    doi = "10.1103/PhysRevC.92.064001",
    journal = "Phys. Rev. C",
    volume = "92",
    number = "6",
    pages = "064001",
    year = "2015"
}

@article{Epelbaum:2019zqc,
    author = "Epelbaum, E. and others",
    title = "{Towards high-order calculations of three-nucleon scattering in chiral effective field theory}",
    eprint = "1907.03608",
    archivePrefix = "arXiv",
    primaryClass = "nucl-th",
    doi = "10.1140/epja/s10050-020-00102-2",
    journal = "Eur. Phys. J. A",
    volume = "56",
    number = "3",
    pages = "92",
    year = "2020"
}

@article{Machleidt:2011zz,
    author = "Machleidt, R. and Entem, D. R.",
    title = "{Chiral effective field theory and nuclear forces}",
    eprint = "1105.2919",
    archivePrefix = "arXiv",
    primaryClass = "nucl-th",
    doi = "10.1016/j.physrep.2011.02.001",
    journal = "Phys. Rept.",
    volume = "503",
    pages = "1--75",
    year = "2011"
}

@article{Ren:2016jna,
    author = "Ren, Xiu-Lei and Li, Kai-Wen and Geng, Li-Sheng and Long, Bing-Wei and Ring, Peter and Meng, Jie",
    title = "{Leading order relativistic chiral nucleon-nucleon interaction}",
    eprint = "1611.08475",
    archivePrefix = "arXiv",
    primaryClass = "nucl-th",
    reportNumber = "CTP-SCU-2016012",
    doi = "10.1088/1674-1137/42/1/014103",
    journal = "Chin. Phys. C",
    volume = "42",
    number = "1",
    pages = "014103",
    year = "2018"
}

@article{Ren:2012aj,
    author = "Ren, X. -L. and Geng, L. S. and Martin Camalich, J. and Meng, J. and Toki, H.",
    title = "{Octet baryon masses in next-to-next-to-next-to-leading order covariant baryon chiral perturbation theory}",
    eprint = "1209.3641",
    archivePrefix = "arXiv",
    primaryClass = "nucl-th",
    doi = "10.1007/JHEP12(2012)073",
    journal = "JHEP",
    volume = "12",
    pages = "073",
    year = "2012"
}

@article{Xiao:2018rvd,
    author = "Xiao, Yang and Ren, Xiu-Lei and Lu, Jun-Xu and Geng, Li-Sheng and Mei\ss{}ner, Ulf-G.",
    title = "{Octet baryon magnetic moments at next-to-next-to-leading order in covariant chiral perturbation theory}",
    eprint = "1803.04251",
    archivePrefix = "arXiv",
    primaryClass = "hep-ph",
    doi = "10.1140/epjc/s10052-018-5960-4",
    journal = "Eur. Phys. J. C",
    volume = "78",
    pages = "489",
    year = "2018"
}

@article{Siemens:2016hdi,
    author = "Siemens, D. and Bernard, V. and Epelbaum, E. and Gasparyan, A. and Krebs, H. and Mei\ss{}ner, Ulf-G.",
    title = "{Elastic pion-nucleon scattering in chiral perturbation theory: A fresh look}",
    eprint = "1602.02640",
    archivePrefix = "arXiv",
    primaryClass = "nucl-th",
    doi = "10.1103/PhysRevC.94.014620",
    journal = "Phys. Rev. C",
    volume = "94",
    number = "1",
    pages = "014620",
    year = "2016"
}

@article{Chen:2012nx,
    author = "Chen, Yun-Hua and Yao, De-Liang and Zheng, H. Q.",
    title = "{Analyses of pion-nucleon elastic scattering amplitudes up to $O(p^4)$ in extended-on-mass-shell subtraction scheme}",
    eprint = "1212.1893",
    archivePrefix = "arXiv",
    primaryClass = "hep-ph",
    doi = "10.1103/PhysRevD.87.054019",
    journal = "Phys. Rev. D",
    volume = "87",
    pages = "054019",
    year = "2013"
}

@article{Xiao:2018jot,
    author = "Xiao, Yang and Geng, Li-Sheng and Ren, Xiu-Lei",
    title = "{Covariant chiral nucleon-nucleon contact Lagrangian up to order $\mathcal{O}(q^4)$}",
    eprint = "1812.03005",
    archivePrefix = "arXiv",
    primaryClass = "nucl-th",
    doi = "10.1103/PhysRevC.99.024004",
    journal = "Phys. Rev. C",
    volume = "99",
    number = "2",
    pages = "024004",
    year = "2019"
}

@article{Stoks:1993tb,
    author = "Stoks, V. G. J. and Klomp, R. A. M. and Rentmeester, M. C. M. and de Swart, J. J.",
    title = "{Partial wave analaysis of all nucleon-nucleon scattering data below 350-MeV}",
    doi = "10.1103/PhysRevC.48.792",
    journal = "Phys. Rev. C",
    volume = "48",
    pages = "792--815",
    year = "1993"
}

@article{Arndt:1994br,
    author = "Arndt, Richard A. and Strakovsky, Igor I. and Workman, Ron L.",
    title = "{An Updated analysis of N N elastic scattering data to 1.6-GeV}",
    eprint = "nucl-th/9407035",
    archivePrefix = "arXiv",
    reportNumber = "VPI-CAPS-7-2",
    doi = "10.1103/PhysRevC.50.2731",
    journal = "Phys. Rev. C",
    volume = "50",
    pages = "2731--2741",
    year = "1994"
}

@article{Xiao:2020ozd,
    author = "Xiao, Yang and Wang, Chun-Xuan and Lu, Jun-Xu and Geng, Li-Sheng",
    title = "{Two-pion exchange contributions to the nucleon-nucleon interaction in covariant baryon chiral perturbation theory}",
    eprint = "2007.13675",
    archivePrefix = "arXiv",
    primaryClass = "nucl-th",
    doi = "10.1103/PhysRevC.102.054001",
    journal = "Phys. Rev. C",
    volume = "102",
    number = "5",
    pages = "054001",
    year = "2020"
}

@article{Wang:2021kos,
    author = "Wang, Chun-Xuan and Lu, Jun-Xu and Xiao, Yang and Geng, Li-Sheng",
    title = "{Nonperturbative two-pion exchange contributions to the nucleon-nucleon interaction in covariant baryon chiral perturbation theory}",
    eprint = "2110.05278",
    archivePrefix = "arXiv",
    primaryClass = "nucl-th",
    doi = "10.1103/PhysRevC.105.014003",
    journal = "Phys. Rev. C",
    volume = "105",
    number = "1",
    pages = "014003",
    year = "2022"
}

@article{Lu:2021gsb,
    author = "Lu, Jun-Xu and Wang, Chun-Xuan and Xiao, Yang and Geng, Li-Sheng and Meng, Jie and Ring, Peter",
    title = "{Accurate Relativistic Chiral Nucleon-Nucleon Interaction up to Next-to-Next-to-Leading Order}",
    eprint = "2111.07766",
    archivePrefix = "arXiv",
    primaryClass = "nucl-th",
    doi = "10.1103/PhysRevLett.128.142002",
    journal = "Phys. Rev. Lett.",
    volume = "128",
    number = "14",
    pages = "142002",
    year = "2022"
}

@article{Geng:2008mf,
    author = "Geng, L. S. and Martin Camalich, J. and Alvarez-Ruso, L. and Vicente Vacas, M. J.",
    title = "{Leading SU(3)-breaking corrections to the baryon magnetic moments in Chiral Perturbation Theory}",
    eprint = "0805.1419",
    archivePrefix = "arXiv",
    primaryClass = "hep-ph",
    doi = "10.1103/PhysRevLett.101.222002",
    journal = "Phys. Rev. Lett.",
    volume = "101",
    pages = "222002",
    year = "2008"
}

@article{Shi:2018rhk,
    author = "Shi, Rui-Xiang and Xiao, Yang and Geng, Li-Sheng",
    title = "{Magnetic moments of the spin-1/2 singly charmed baryons in covariant baryon chiral perturbation theory}",
    eprint = "1812.07833",
    archivePrefix = "arXiv",
    primaryClass = "hep-ph",
    doi = "10.1103/PhysRevD.100.054019",
    journal = "Phys. Rev. D",
    volume = "100",
    number = "5",
    pages = "054019",
    year = "2019"
}

@article{Liu:2018euh,
    author = "Liu, Ming-Zhu and Xiao, Yang and Geng, Li-Sheng",
    title = "{Magnetic moments of the spin-1/2 doubly charmed baryons in covariant baryon chiral perturbation theory}",
    eprint = "1807.00912",
    archivePrefix = "arXiv",
    primaryClass = "hep-ph",
    doi = "10.1103/PhysRevD.98.014040",
    journal = "Phys. Rev. D",
    volume = "98",
    number = "1",
    pages = "014040",
    year = "2018"
}

@article{Ren:2016aeo,
    author = "Ren, Xiu-Lei and Alvarez-Ruso, L. and Geng, Li-Sheng and Ledwig, T. and Meng, Jie and Vicente Vacas, M. J.",
    title = "{Consistency between SU(3) and SU(2) covariant baryon chiral perturbation theory for the nucleon mass}",
    eprint = "1606.03820",
    archivePrefix = "arXiv",
    primaryClass = "nucl-th",
    doi = "10.1016/j.physletb.2017.01.024",
    journal = "Phys. Lett. B",
    volume = "766",
    pages = "325--333",
    year = "2017"
}

@article{Ren:2014vea,
    author = "Ren, Xiu-Lei and Geng, Li-Sheng and Meng, Jie",
    title = "{Scalar strangeness content of the nucleon and baryon sigma terms}",
    eprint = "1404.4799",
    archivePrefix = "arXiv",
    primaryClass = "hep-ph",
    doi = "10.1103/PhysRevD.91.051502",
    journal = "Phys. Rev. D",
    volume = "91",
    number = "5",
    pages = "051502",
    year = "2015"
}

@article{Ren:2013wxa,
    author = "Ren, Xiu-Lei and Geng, Li-Sheng and Meng, Jie",
    title = "{Baryon chiral perturbation theory with Wilson fermions up to $\mathcal{O}(a^2)$ and discretization effects of latest $n_f=2+1$ LQCD octet baryon masses}",
    eprint = "1311.7234",
    archivePrefix = "arXiv",
    primaryClass = "hep-ph",
    doi = "10.1140/epjc/s10052-014-2754-1",
    journal = "Eur. Phys. J. C",
    volume = "74",
    number = "2",
    pages = "2754",
    year = "2014"
}

@article{Ren:2013oaa,
    author = "Ren, Xiu-Lei and Geng, Li-Sheng and Meng, Jie",
    title = "{Decuplet baryon masses in covariant baryon chiral perturbation theory}",
    eprint = "1307.1896",
    archivePrefix = "arXiv",
    primaryClass = "nucl-th",
    doi = "10.1103/PhysRevD.89.054034",
    journal = "Phys. Rev. D",
    volume = "89",
    number = "5",
    pages = "054034",
    year = "2014"
}

@article{Ren:2013dzt,
    author = "Ren, Xiu-Lei and Geng, Lisheng and Meng, Jie and Toki, Hiroshi",
    title = "{Virtual decuplet effects on octet baryon masses in covariant baryon chiral perturbation theory}",
    eprint = "1302.1953",
    archivePrefix = "arXiv",
    primaryClass = "nucl-th",
    doi = "10.1103/PhysRevD.87.074001",
    journal = "Phys. Rev. D",
    volume = "87",
    number = "7",
    pages = "074001",
    year = "2013"
}

@article{Machleidt:2024bwl,
    author = "Machleidt, Ruprecht and Sammarruca, Francesca",
    title = "{Recent advances in chiral EFT based nuclear forces and their applications}",
    eprint = "2402.14032",
    archivePrefix = "arXiv",
    primaryClass = "nucl-th",
    month = "2",
    year = "2024"
}

@article{Weinberg:1991um,
    author = "Weinberg, Steven",
    title = "{Effective chiral Lagrangians for nucleon - pion interactions and nuclear forces}",
    reportNumber = "UTTG-03-91",
    doi = "10.1016/0550-3213(91)90231-L",
    journal = "Nucl. Phys. B",
    volume = "363",
    pages = "3--18",
    year = "1991"
}

@article{Weinberg:1992yk,
    author = "Weinberg, Steven",
    title = "{Three body interactions among nucleons and pions}",
    eprint = "hep-ph/9209257",
    archivePrefix = "arXiv",
    reportNumber = "UTTG-11-92",
    doi = "10.1016/0370-2693(92)90099-P",
    journal = "Phys. Lett. B",
    volume = "295",
    pages = "114--121",
    year = "1992"
}

@article{RodriguezEntem:2020jgp,
    author = "Rodriguez Entem, David and Machleidt, Ruprecht and Nosyk, Yevgen",
    title = "{Nucleon-Nucleon Scattering Up to N$^5$LO in Chiral Effective Field Theory}",
    doi = "10.3389/fphy.2020.00057",
    journal = "Front. in Phys.",
    volume = "8",
    pages = "57",
    year = "2020"
}

@article{Shi:2021kmm,
    author = "Shi, Rui-Xiang and Geng, Li-Sheng",
    title = "{Magnetic moments of the spin-$\frac{3}{2}$ doubly charmed baryons in covariant baryon chiral perturbation theory}",
    eprint = "2103.07260",
    archivePrefix = "arXiv",
    primaryClass = "hep-ph",
    doi = "10.1103/PhysRevD.103.114004",
    journal = "Phys. Rev. D",
    volume = "103",
    number = "11",
    pages = "114004",
    year = "2021"
}

@article{Lu:2025syk,
    author = "Lu, Jun-Xu and Xiao, Yang and Liu, Zhi-Wei and Geng, Li-Sheng",
    title = "{Relativistic chiral nuclear forces: Status and prospects}",
    eprint = "2501.17185",
    archivePrefix = "arXiv",
    primaryClass = "nucl-th",
    doi = "10.1142/S0218301325430074",
    journal = "Int. J. Mod. Phys. E",
    volume = "34",
    number = "11",
    pages = "2543007",
    year = "2025"
}

@article{Fettes:2000gb,
    author = "Fettes, Nadia and Meissner, Ulf-G. and Mojzis, Martin and Steininger, Sven",
    title = "{The Chiral effective pion nucleon Lagrangian of order p**4}",
    eprint = "hep-ph/0001308",
    archivePrefix = "arXiv",
    reportNumber = "FZJ-IKP-TH-2000-04",
    doi = "10.1006/aphy.2000.6059",
    journal = "Annals Phys.",
    volume = "283",
    pages = "273--302",
    year = "2000",
    note = "[Erratum: Annals Phys. 288, 249--250 (2001)]"
}

@article{UCNA:2010les,
    author = "Liu, J. and others",
    collaboration = "UCNA",
    title = "{Determination of the Axial-Vector Weak Coupling Constant with Ultracold Neutrons}",
    eprint = "1007.3790",
    archivePrefix = "arXiv",
    primaryClass = "nucl-ex",
    doi = "10.1103/PhysRevLett.105.181803",
    journal = "Phys. Rev. Lett.",
    volume = "105",
    pages = "181803",
    year = "2010"
}

@article{Kaiser:1999ff,
    author = "Kaiser, Norbert",
    title = "{Chiral 3 pi exchange N N potentials: Results for representation invariant classes of diagrams}",
    eprint = "nucl-th/9910044",
    archivePrefix = "arXiv",
    doi = "10.1103/PhysRevC.61.014003",
    journal = "Phys. Rev. C",
    volume = "61",
    pages = "014003",
    year = "2000"
}

@article{Kaiser:1999jg,
    author = "Kaiser, Norbert",
    title = "{Chiral three pi exchange N N potentials: Results for diagrams proportional to g(A)**4 and g(A)**6}",
    eprint = "nucl-th/9912054",
    archivePrefix = "arXiv",
    doi = "10.1103/PhysRevC.62.024001",
    journal = "Phys. Rev. C",
    volume = "62",
    pages = "024001",
    year = "2000"
}

@article{Liu:2017jxz,
    author = "Liu, Xiao and Ma, Yan-Qing and Wang, Chen-Yu",
    title = "{A Systematic and Efficient Method to Compute Multi-loop Master Integrals}",
    eprint = "1711.09572",
    archivePrefix = "arXiv",
    primaryClass = "hep-ph",
    doi = "10.1016/j.physletb.2018.02.026",
    journal = "Phys. Lett. B",
    volume = "779",
    pages = "353--357",
    year = "2018"
}

@article{Liu:2021wks,
    author = "Liu, Xiao and Ma, Yan-Qing",
    title = "{Multiloop corrections for collider processes using auxiliary mass flow}",
    eprint = "2107.01864",
    archivePrefix = "arXiv",
    primaryClass = "hep-ph",
    doi = "10.1103/PhysRevD.105.L051503",
    journal = "Phys. Rev. D",
    volume = "105",
    number = "5",
    pages = "L051503",
    year = "2022"
}

@article{Liu:2022chg,
    author = "Liu, Xiao and Ma, Yan-Qing",
    title = "{AMFlow: A Mathematica package for Feynman integrals computation via auxiliary mass flow}",
    eprint = "2201.11669",
    archivePrefix = "arXiv",
    primaryClass = "hep-ph",
    doi = "10.1016/j.cpc.2022.108565",
    journal = "Comput. Phys. Commun.",
    volume = "283",
    pages = "108565",
    year = "2023"
}

@article{Liu:2022mfb,
    author = "Liu, Zhi-Feng and Ma, Yan-Qing",
    title = "{Determining Feynman Integrals with Only Input from Linear Algebra}",
    eprint = "2201.11637",
    archivePrefix = "arXiv",
    primaryClass = "hep-ph",
    doi = "10.1103/PhysRevLett.129.222001",
    journal = "Phys. Rev. Lett.",
    volume = "129",
    number = "22",
    pages = "222001",
    year = "2022"
}

@article{Liu:2020kpc,
    author = "Liu, Xiao and Ma, Yan-Qing and Tao, Wei and Zhang, Peng",
    title = "{Calculation of Feynman loop integration and phase-space integration via auxiliary mass flow}",
    eprint = "2009.07987",
    archivePrefix = "arXiv",
    primaryClass = "hep-ph",
    doi = "10.1088/1674-1137/abc538",
    journal = "Chin. Phys. C",
    volume = "45",
    number = "1",
    pages = "013115",
    year = "2021"
}

@article{Liu:2022tji,
    author = "Liu, Zhi-Feng and Ma, Yan-Qing",
    title = "{Automatic computation of Feynman integrals containing linear propagators via auxiliary mass flow}",
    eprint = "2201.11636",
    archivePrefix = "arXiv",
    primaryClass = "hep-ph",
    doi = "10.1103/PhysRevD.105.074003",
    journal = "Phys. Rev. D",
    volume = "105",
    number = "7",
    pages = "074003",
    year = "2022"
}

@article{Liang:2025adz,
    author = "Liang, Ze-Rui and Chen, Han-Xue and Guo, Feng-Kun and Guo, Zhi-Hui and Yao, De-Liang",
    title = "{Two-Loop Extraction of the Pion-Nucleon Sigma Term}",
    eprint = "2508.11435",
    archivePrefix = "arXiv",
    primaryClass = "hep-ph",
    month = "8",
    year = "2025"
}

@article{Liang:2025cjd,
    author = "Liang, Ze-Rui and Chen, Han-Xue and Guo, Feng-Kun and Guo, Zhi-Hui and Yao, De-Liang",
    title = "{Chiral representation of the nucleon mass at leading two-loop order}",
    eprint = "2502.19168",
    archivePrefix = "arXiv",
    primaryClass = "hep-ph",
    doi = "10.1007/JHEP04(2025)192",
    journal = "JHEP",
    volume = "04",
    pages = "192",
    year = "2025"
}

@article{Chemtob:1972xt,
    author = "Chemtob, M. and Durso, J. W. and Riska, D. O.",
    title = "{Two-pion-exchange nucleon-nucleon potential}",
    doi = "10.1016/0550-3213(72)90345-8",
    journal = "Nucl. Phys. B",
    volume = "38",
    pages = "141--206",
    year = "1972"
}

@article{Lu:2018zof,
    author = "Lu, Jung-Xu and Geng, Li-Sheng and Ren, Xiu-Lei and Du, Meng-Lin",
    title = "{Meson-baryon scattering up to the next-to-next-to-leading order in covariant baryon chiral perturbation theory}",
    eprint = "1812.03799",
    archivePrefix = "arXiv",
    primaryClass = "nucl-th",
    doi = "10.1103/PhysRevD.99.054024",
    journal = "Phys. Rev. D",
    volume = "99",
    number = "5",
    pages = "054024",
    year = "2019"
}

@article{Arndt:2007qn,
    author = "Arndt, R. A. and Briscoe, W. J. and Strakovsky, I. I. and Workman, R. L.",
    title = "{Updated analysis of NN elastic scattering to 3-GeV}",
    eprint = "0706.2195",
    archivePrefix = "arXiv",
    primaryClass = "nucl-th",
    doi = "10.1103/PhysRevC.76.025209",
    journal = "Phys. Rev. C",
    volume = "76",
    pages = "025209",
    year = "2007"
}

@article{Siemens:2016jwj,
    author = "Siemens, D. and Ruiz de Elvira, J. and Epelbaum, E. and Hoferichter, M. and Krebs, H. and Kubis, B. and Mei{\ss}ner, U. -G.",
    title = "{Reconciling threshold and subthreshold expansions for pion{\textendash}nucleon scattering}",
    eprint = "1610.08978",
    archivePrefix = "arXiv",
    primaryClass = "nucl-th",
    reportNumber = "INT-PUB-16-035, NSF-KITP-16-158",
    doi = "10.1016/j.physletb.2017.04.039",
    journal = "Phys. Lett. B",
    volume = "770",
    pages = "27--34",
    year = "2017"
}

\end{document}